\documentclass[submit]{smj}
\pdfoutput=1  

\usepackage{setspace,xcolor,bm,multirow,multicol}

\graphicspath{{./figures/}}  

\Author{Lizandra C. Fabio\Affil{1}, 
        Vanessa Barros\Affil{1}, 
				Cristian Villegas\Affil{2},
        \and Jalmar M. F. Carrasco \Affil{1}
}
\AuthorRunning{Lizandra C. Fabio \textrm{et al.}}

\Affiliations{

\item Federal University of Bahia, 
      Brazil

\item University of Sao Paulo,
      Brazil

}   

\CorrAddress{Jalmar M. F. Carrasco, 
             Department of Statistics, 
             Federal University of Bahia,
             Brazil}
\CorrEmail{carrascojalmar@gmail.com}
\CorrPhone{(+55)\;71\;9910\;45055}

\Title{A novel multivariate regression model for unbalanced binary data : a strong conjugacy under random effect approach}
\TitleRunning{ }

\Abstract{
In this paper, we deduce a new multivariate regression model designed to fit correlated binary data. The multivariate distribution is derived from a Bernoulli mixed model with a nonnormal random intercept on the marginal approach. The random effect distribution is assumed to be the generalized log-gamma (GLG) distribution by considering a particular parameter setting. The complement log-log function is specified to lead to strong conjugacy between the response variable and random effect. The new discrete multivariate distribution, named MBerGLG distribution, has location and dispersion parameters. The MBerGLG distribution leads to the MBerGLG regression (MBerGLGR) model, providing an alternative approach to fitting both unbalanced and balanced correlated response binary data. Monte Carlo simulation studies show that its maximum likelihood estimators are unbiased, efficient, and consistent asymptotically.  The randomized quantile residuals are performed to identify possible departures from the proposal model and the data and detect atypical subjects. Finally, two applications are presented in the data analysis section.
}


\begin{document}
\maketitle

\section{Introduction}
\label{sec:Intro}

Binary responses with repeated measurements are often modeled by generalized linear mixed models (GLMM). The hierarchical structure of the GLMM is a natural extension of the generalized linear model (GLM) where the response variable belongs to the exponential family of distributions (\citealp{Diggle04, Hilbe2016}). Random effects are included in the systematic component to induce intraclass correlation and to draw inferences on it (\citealp{Breslow93, Geert2010}). The random effect approach has been used in longitudinal analysis where the study aims to investigate the behavior of the subject over time (\citealp{Molenberghs10, Mol12, Lee01}). 
GLMM are also useful in studies where the interest lies in the evolution of marginal population averages and in multivariate regression models (\citealp{Fabio12, Fabio22}). In the literature, the Bahadur model (\citealp{Bahadur1961}), the multivariate Probit model (\citealp{Mol12}), and the Dale model (\citealp{Dale1986}) have been suggested for analyzing correlated binary data.  Multivariate regression models can also be derived from the random effect approach as an alternative to existing models, for example, the Probit-normal models (\citealp{Molenberghs10}). Based on the marginal approach, the multivariate Binomial negative regression and the multivariate inverted Dirichlet distribution extend regression model were deduced by \citealp{Fabio12} and \citealp{Fabio22}, respectively.
Following this flexible approach to deduce new multivariate marginal functions, we propose deriving a multivariate model from a Bernoulli mixed model with a non-normally distributed random intercept. We assume the generalized log-gamma density function by specifying a particular parameter setting for the random effect. We demonstrate a strong conjugacy between the response variable response and random effect distributions by considering the complementary log-log link function. 
The MBerGLG distribution is performed analytically and has an explicit form. The MBerGLG distribution leads to the MBerGLG regression (MBerGLGR) model. The advantage of the MBerGLGR model over existing multivariate models for binary response data is that its moments and intraclass correlation can be expressed analytically. Additionally, its marginal probability function resembles the Bernoulli model's, and its inferential process is not cumbersome. The maximum likelihood estimates are computed by using the quasi-Newton ({\tt  Broyden-Fletcher-Goldfarb-Shanno (BFGS)}) method that is available in $optim(\cdot, method=``BFGS")$ function of the \texttt{R} software. Estimation equations based on score functions are calculated with respect to parameters as an alternative to the inferential process, as well as the expressions of the components of the observed Fisher information to obtain interval estimation and hypothesis tests on the model parameters. Monte Carlo simulations are performed to evaluate the asymptotic behavior of the estimators. The simulation results also show that the MBerGLGR models fit unbalanced and balanced data appropriately. The randomized quantile residuals analysis is suggested to identify potential departures from the proposal model and to detect atypical subjects. 

The remainder of this paper is organized as follows: Section \ref{GLG} provides an overview of the GLG distribution. In section \ref{BerGLG}, we derive the MBerGLG distribution from the random intercept Bernoulli-GLG models and present its moments and inferential procedures. Section \ref{sec:simula} contains simulation studies to evaluate the asymptotic behavior of the maximum likelihood estimators. In section \ref{sec:residuos}, we propose the randomized quantile residuals for the MBerGLGR model, and in section \ref{sec:apli}, we present two applications to illustrate the proposed methodology.

\section{Generalized log-gamma distribution}
\label{GLG}

\begin{figure}
\centering
\resizebox*{4.5cm}{!}{\includegraphics{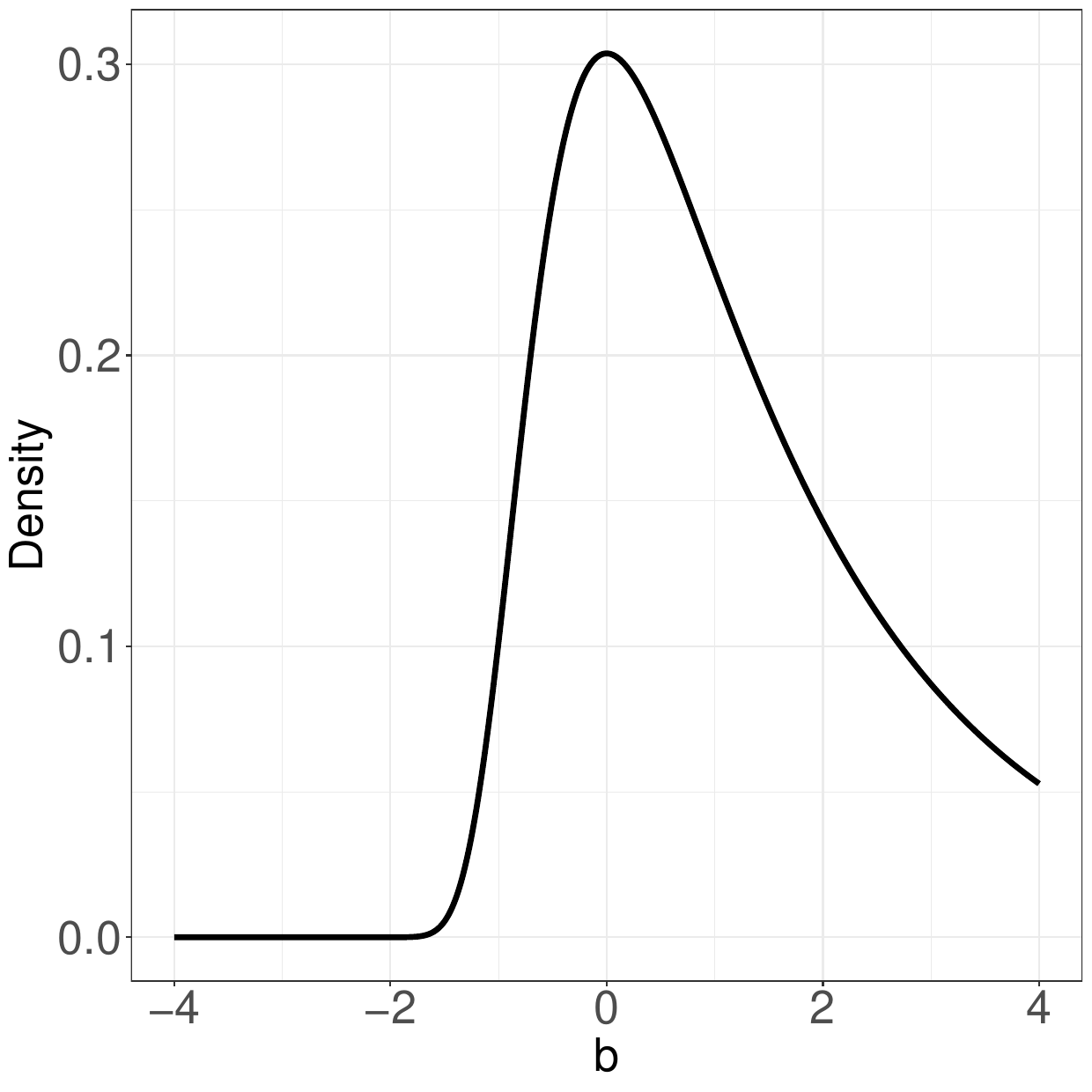}}
\resizebox*{4.5cm}{!}{\includegraphics{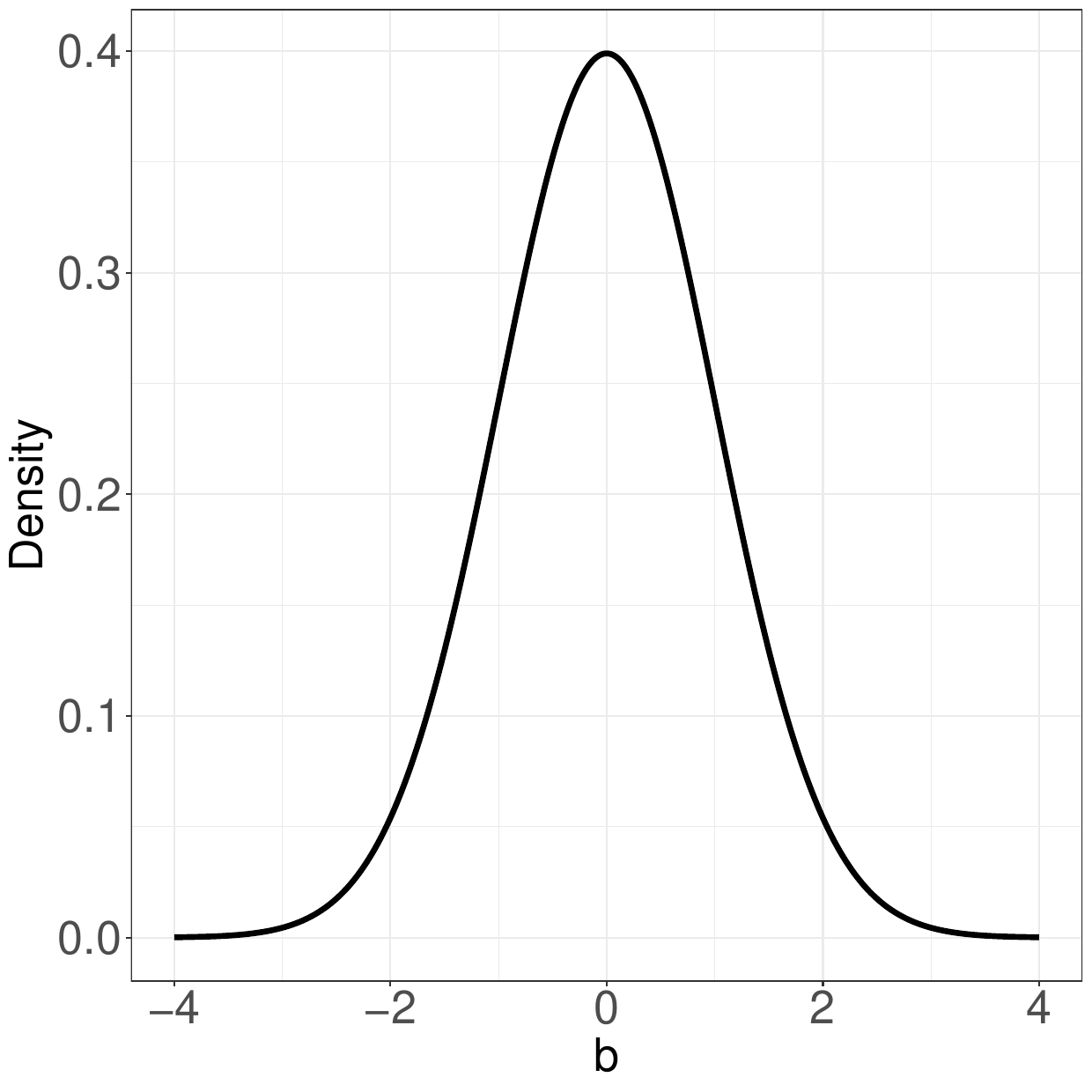}}
\resizebox*{4.5cm}{!}{\includegraphics{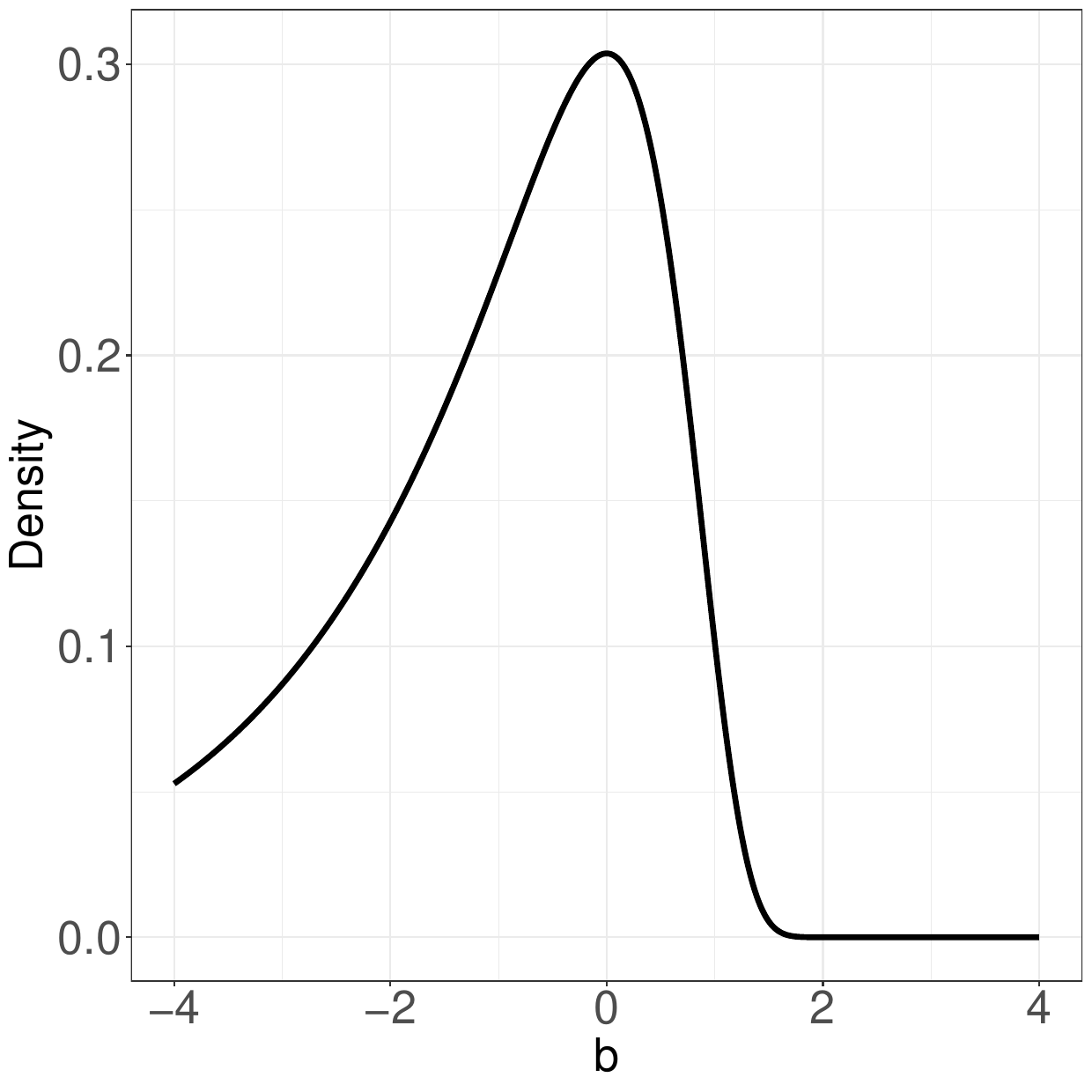}}
\caption{The GLG density function for the parameters $\mu=0$, $\sigma=1$, $\lambda=-2$ (left), $\lambda=0$ (center) and $\lambda=2$ (right).} 
\label{densi}
\end{figure}

The Generalized log-gamma distribution was suggested by \citealp{Lawless80}, and it has been widely applied in the areas of survival analysis and reliability (\citealp{Chien2004, Cox2007, Ortega2009} ). Let $b\in \mathbb{R}$ be a random variable following a generalized log-gamma (GLG) distribution, with its probability density function (pdf) given by the following expression:
\begin{align} 
\label{fb} 
f(b; \mu, \sigma, \lambda)=
  \begin{cases} \frac{c(\lambda)}{\sigma}
    \exp\left [\frac{(b - \mu)}{\lambda \sigma} - \frac{1}{\lambda^2}
      \exp \left \{ \frac{\lambda(b - \mu)}{ \sigma} \right \} \right ],
    & \textrm{if} \quad  \lambda \ne  0, \\
    \frac{1}{\sigma\sqrt{2\pi}}\exp\left \{-\frac{(b - \mu)^2}{2\sigma^2}
    \right \}, &\textrm{if} \quad \lambda=0,
  \end{cases}
\end{align}
where $\mu \in \mathbb{R}$, $\sigma > 0$, and $\lambda \in \mathbb{R}$ represent the location, scale, and shape parameters, respectively. Moreover, we have $c(\lambda) = |\lambda|(\lambda^{-2})^{\lambda^{-2}}/\Gamma(\lambda^{-2})$, where $\Gamma(\cdot)$ denotes the gamma function. We denoted $b \sim {\rm GLG}(\mu, \sigma, \lambda)$.
When $\lambda=1$, the expression (\ref{fb}) becomes the extreme value distribution. For $\lambda < 0$ and $\lambda > 0$
the GLG pdf of $b$ is skewed to the right and left, respectively. Finally, at $\lambda = 0$, the pdf in (\ref{fb}) simplifies to the normal distribution. Figure~\ref{densi} illustrates the GLG density function's behavior with respect to the $\lambda$ parameter. 
We have for $\lambda \ne 0$ the following moments:
$$
{\rm E}(b) = \mu + \sigma \left \{ \frac{\psi(\lambda^{-2}) + \log
    \lambda^{-2}} {|\lambda|} \right \} \ \ \ \ {\rm and} \ \ \ \ {\rm
  Var}(b) = \sigma^2 \frac{\psi'(\lambda^{-2})}{\lambda^2},
$$
where $\psi(\cdot)$ and $\psi'(\cdot)$ denote, respectively, the digamma and trigamma functions. For $\lambda = 0$ we have the normal case for which ${\rm E}(b) = \mu$ and ${\rm Var}(b) = \sigma^2$.

\section{The multivariate BerGLG regression model}
\label{BerGLG}

Let $y_{ij}$ represent the $j$th measurement taken on the $i$th subject or cluster, where $j=1,\ldots,m_i$ and $i=1,\ldots,n$. Let $b_i$ be a random effect following the GLG distribution \citep{Lawless80}. Following \cite{Fabio12} and \cite{Fabio22}, we assuming that $y_{ij}|b_i$ are independent outcomes with a probability mass function (pmf) described by a Bernoulli distribution, we propose a random intercept Bernoulli-GLG model with the following hierarchical structure:
\begin{eqnarray*}
&(i)&  \quad y_{ij} |b_i \stackrel{\rm ind} {\sim}  {\rm Bernoulli}(u_{ij}),\\			
&(ii)&  \quad g(u_{ij})  =   \log(-\log(1 - u_{ij})) = \bm x_{ij}^{\top}\bm \beta + b_i\\ 
&(iii)&  \quad b_i  \stackrel{\rm i.i.d} {\sim} {\rm GLG}(0, \lambda, \lambda),
\end{eqnarray*}
where $\bm{x}_{ij} = (x_{ij1}, \ldots, x_{ijp})^{\top}$ contains the values of explanatory variables, $\bm{\beta} = (\beta_1, \ldots, \beta_p)^{\top}$ is the parameter vector of the systematic component and $g(\cdot)$ the link function. If $\lambda=0$, the random intercept Bernoulli-GLG model simplifies to the random intercept Bernoulli-normal model \citep{Breslow93}. We denote $f(y_{ij}|{ b_i}; \bm{\beta})$ as the probability mass function (pmf) of $y_{ij}|{ b_i}$ and $f({b_i}; \lambda)$ as the probability density function (pdf) of $b_i$. The marginal pmf of $\bm{y}_i = (y_{i1},\ldots,y_{im_i})^{\top}$ with $\bm{\theta} = (\bm{\beta}^{\top}, \lambda)^{\top}$ is given by
\begin{equation} 
\label{e2} 
f(\bm{y}_i ; \bm{\theta})= \int_{-\infty}^{+\infty} \Big\{\prod_{j=1}^{m_i}f(y_{ij}|{b_i}; \bm \beta)\Big\}f({b_i}; \lambda)d{b_i},
\end{equation}
and $f(\bm{y};\bm{\theta})=\prod_{i=1}^{n}f(\bm{y}_i;\bm{\theta})$ where $\bm{y}=(\bm{y}_1,\ldots,\bm{y}_{n})^{\top}$. The marginal pmf (\ref{e2}) and the log-likelihood function $\ell(\bm{ \theta}) =\sum_{i=1}^{n}\log( f(\bm{y}_{i};\bm{\theta}))$ do not have explicit form and numerical methods are required to solve the integral and maximize the approximated $\ell(\bm{ \theta})$ function. The marginal (\ref{e2}) has an explicit form for the proposed hierarchical model due to the strong conjugacy between the response variable and the random effect, assuming the complement log-log link function. In this case, the marginal probability mass function (pmf) of $\bm{y}_i$ takes the following form: 
\begin{eqnarray}  
\label{eqG1}
f(\bm y_i; \bm{\theta})&=& \frac{\phi^{\phi}}{\Gamma(\phi)} \int_{-\infty}^{+\infty}e^{e^{b_i}({\sum_{j=1}^{m_i} \mu_{ij}y_{ij}-\mu_{i+}
)}}e^{\phi b_i}e^{-\phi e^{b_i}}\prod_{j=1}^{m_i}\Bigg(1- e^{-\mu_{ij}e^b_i} \Bigg)^{y_{ij}} db_i,
\end{eqnarray}
where $\bm{\theta}=(\bm{\beta}^{\top}, \phi)^{\top}$, with $\phi=\lambda^{-2}$, $\mu_{i+}=\sum_{j=1}^{m_i}\mu_{ij}$ and $\mu_{ij}=\exp(\bm{x}_{ij}^{\top}\beta)$, for $j=1,\ldots, m_i$ and $i=1,\ldots,n$. Considering the variable transformation $t_i = \exp(b_i)$ and performing some algebraic manipulations, expression (\ref{eqG1}) can be rewritten as 
\begin{eqnarray}  
\label{eqG2}
f(\bm y_i; \bm{\theta})&=& \frac{\phi^{\phi}}{\Gamma(\phi)} \int_{0}^{+\infty}   e^{-t_i(\mu_{i+}-\sum_{j=1}^{m_i}\mu_{ij}y_{ij}+\phi)}t_i^{\phi-1}\prod_{j=1}^{m_i}\Bigg(1- e^{-\mu_{ij}t_i}\Bigg)^{y_{ij}}dt_i.
\end{eqnarray}
By employing the binomial theorem to expand the term $(1- e^{-\mu_{ij}t_i})^{y_{ij}}$ we can express equation (\ref{eqG2}) as follows
\begin{eqnarray}  
\label{eqG3}
f(\bm y_i; \bm{\theta})&=&\frac{\phi^{\phi}}{\Gamma(\phi)} \int_{0}^{+\infty}
   e^{-t_i(\mu_{i+}-\sum_{j=1}^{m_i}\mu_{ij}y_{ij}+\phi)}t_i^{\phi-1}\prod_{j=1}^{m_i}
\Bigg(\sum_{k=0}^{y_{ij}}(-1)^k e^{-\mu_{ij}t_ik}\Bigg)dt_i.
\end{eqnarray}
Now we notice that expanding the  product of sums $\prod_{j=1}^{m_i}
\Big(\sum_{k=0}^{y_{ij}}(-1)^k \exp(-\mu_{ij}t_ik)\Big)$ will result in a sum of  $2^{\sum_{j=1}^{m_i} y_{ij}}$
terms of the type $(-1)^{\sum_{j=1}^{m_i}k_j} \exp(-t_i(\sum_{j=1}^{m_i}\mu_{ij}k_j)),$ where $k_j$ is either $0$ or $1$.
Given $i \in \{1,\cdots,n\}$ we consider the following set 
$\widetilde{K}=\{ \alpha=(k_1,\cdots,k_{m_i}); \hspace{0.2cm} 0\leq k_j \leq y_{ij},\hspace{0.2cm} k_j \in \{0,1\},\hspace{0.2cm} j =1,\cdots,m_i\}$. Note that $\widetilde{K}$ could be interpreted as a matrix with $2^{\sum_{j=1}^{m_i} y_{ij}}$ rows and $m_i$ columns with its elements being $0$'s and 1's.
So  denoting $\sum_{j=1}^{m_i}k_j$ as $k_+$  we can rewrite expression \eqref{eqG3} as follows
\begin{eqnarray}  
\label{eqG4}
f(\bm y_i; \bm{\theta})&=& \frac{\phi^{\phi}}{\Gamma(\phi)} \sum_{\alpha \in \widetilde{K}}\int_{0}^{+\infty}
   e^{-t_i(\mu_{i+}-\sum_{j=1}^{m_i}\mu_{ij}y_{ij}+\phi)}t_i^{\phi-1}(-1)^{k_+} e^{-t_i(\sum_{j=1}^{m_i}\mu_{ij}k_j)}dt_i,
\end{eqnarray}
where the integral is the probability density function gamma with parameters $\phi$ and $(\mu_{i+}+\phi+ \sum_{j=1}^{m_i}\mu_{ij}(k_j-y_{ij})).$ 
Hence, the marginal multivariate model of $\bm{y}_i$ is given by
\begin{eqnarray}  
\label{eqG5}
f(\bm y_i; \bm{\theta})&=&\phi^{\phi} \sum_{\alpha \in \widetilde{K}}
\left \{
(-1)^{k_+} 
\left(  \mu_{i+}+\phi+ \sum_{j=1}^{m_i}\mu_{ij}(k_j-y_{ij})\right)^{-\phi}
\right \}.
\end{eqnarray}
We refer to the probability mass function (pmf) in (\ref{eqG5}) as the Multivariate BerGLG (MBerGLG) distribution. The $m_i$-dimensional random vector with the pmf given in (\ref{eqG5}) is denoted by $\bm{y}_i \sim \text{BerGLG}(\bm{\mu}_i, \phi)$, where $\bm{\mu}_i = (\mu_{i1}, \ldots, \mu_{i m_i})^{\top}$ represents the location parameter vector. The marginal probability functions of $y_{ij}$ is deduced from (\ref{eqG5}) by considering $m_i =1$ and its expression has the following form
\begin{eqnarray}
\label{f10}
f(y_{ij}, \mu_{ij},\phi) = 
 \Bigg \{
\begin{array}{ccc}
     [\phi/(\phi+\mu_{ij})]^{\phi}   & if & y_{ij}=0,  \\
     1-[\phi/(\phi+\mu_{ij})]^{\phi} & if & y_{ij}=1,  
\end{array}
\end{eqnarray}
where the probability of success is the complement of the probability of failure, which is defined as a function of the parameters $\phi$ and $\mu_{ij}$. The moments of $\bm{y}_i \sim \text{BerGLG}(\bm{\mu}_i, \phi)$ are obtained from the  random intercept Bernoulli-GLG model $(i)$-$(iii)$.
\begin{eqnarray}
\label{esp}
 {\rm E}(y_{ij}) &=& {\rm E}[{\rm E}(y_{ij}|b_i)] = 1 - \frac{\phi^{\phi}}{(\mu_{ij} + \phi)^{\phi}},\\   
 {\rm Var}(y_{ij}) &=& {\rm E}[{\rm Var}(y_{ij}|b_i)]  + {\rm Var}[{\rm E}(y_{ij}|b_i)] \nonumber \\
                  &=&\frac{\phi^{\phi}}{(\mu_{ij} + \phi)^{\phi}} + \frac{\phi^{\phi}}{(2\mu_{ij} + \phi)^{\phi}} -2\frac{\phi^{2\phi}}{(\mu_{ij} + \phi)^{2\phi}} \quad \text{and} \nonumber,\\   
{\rm Cov}(y_{ij}, y_{ij'}) &=& {\rm Cov}({\rm E}[y_{ij}|b_i], {\rm E}[y_{ij'}|b_i]) + {\rm E}({\rm Cov}[y_{ij}|b_i], {\rm Cov}[y_{ij'}|b_i]) \nonumber \\
&=&\frac{\phi^{2\phi}}{(\mu_{ij} + \phi)^{2\phi}}-\frac{\phi^{\phi}}{(2\mu_{ij} + \phi)^{\phi}} \nonumber.    
\end{eqnarray}
Further, $\lim_{\phi \rightarrow 0^{+}}{\rm Var}(y_{ij})\rightarrow 0$ and $\sup_{\mu_{ij}\in (0,1)}\lim_{\phi \rightarrow \infty}{\rm Var}(y_{ij})=0.25$ is achieved when $\mu_{ij}=0.6931457$. Based on this fact, the parameter $\phi$ is also considered a precision parameter. For the marginal probability function the expected value of $y_{ij}$ is calculated from (\ref{esp}) and the its variance is given by  Var$(y_{ij}) = \frac{\phi^{\phi}}{(\mu_{ij} + \phi)^{\phi}}\bigg(1-\frac{\phi^{\phi}}{(\mu_{ij} + \phi)^{\phi}}\bigg).$  

\subsection{Estimation}
\label{subsec:like}

Let $\bm{y}_i=(y_{i1},\ldots,y_{im_i})^{\top}$ a vector containing all the binary outcomes for the $i$th subject. The log-likelihood function is then given by
$\ell(\bm{ \theta})=\sum_{i=1}^{n}\{\phi\log{\phi}+\log A_i(\bm{ \theta})\}$,
where 
$A_i(\bm{\theta})=\sum_{\alpha \in \widetilde{K}}\{(-1)^{k_{+}} D_i(\bm{\theta})^{-\phi}\}$, 
and $D_i(\bm{\theta})=\mu_{i+}+\phi+ \sum_{j=1}^{m_i}\mu_{ij}(k_j-y_{ij})$.
The maximum likelihood (ML) estimates $\widehat{\bm \theta}$ of $\bm \theta$ are computed by using the quasi-Newton ({\tt BFGS}) method. Alternatively, we can solve the nonlinear equation obtained by setting the components of the score vector equal to zero, that is, $\bm{U}_{\theta}=(U_{\bm{\beta}}^{\top}, U_{\phi})^{\top}=\bm{0}.$ The score function with respect to the parameter $\bm{\beta}$ is given as
\begin{equation*} 
\label{partialderivativebeta}
U_{\bm{\beta}}(\bm{\theta}) = 
\frac{\partial \ell(\bm{\theta})}{\partial \bm{\beta}} = 
\left( \frac{\partial \ell(\bm{\theta})}{\partial \beta_1}, \ldots, 
\frac{\partial \ell(\bm{\theta})}{\partial \beta_p} \right)^{\top},
\end{equation*}
where
\begin{equation}
\label{lbeta}
    \frac{\partial \ell(\bm{\theta})}{\partial \beta_q} = \sum_{i=1}^n C_{qi}(\bm{\theta}) A_i(\bm{ \theta})^{-1} \quad q= 1,\ldots,p,
\end{equation}
with
$C_{qi}(\bm{\theta})=\sum_{\alpha \in \widetilde{K}} \frac{(-1)^{k_{+}+1}}{D_i(\bm{ \theta})^{\phi+1}}\phi \sum_{j=1}^{m_i}\left[(1-k_j-y_{ij})e^{x_{ij}^\top \beta}x_{ijq}\right]$. The score function 
with respect to the parameter $\phi$ is given by: 
\begin{equation*} 
\label{derivativeone}
U_{\phi}(\bm{\theta}) = 
\frac{\partial \ell(\bm{\theta})}{\partial \phi} = 
\sum_{i=1}^{n} \left \{ (\log \phi+1) + 
\frac{A_i(\bm{\theta})^{\prime}}{A_i(\bm{ \theta})} \right \},
\end{equation*} 
where 
\begin{equation*} \label{derivativeAi}
A_i(\bm{ \theta})^{\prime}=\frac{\partial A_i(\bm{ \theta})}{\partial \phi}=
\sum_{\alpha \in \widetilde{K}} 
\left \{
(-1)^{k_{+}+1} 
\left(
\frac{\log D_i(\bm{ \theta})}{D_i(\bm{ \theta})^{\phi}}+
\frac{\phi}{D_i(\bm{ \theta})^{\phi+1}}
\right) 
\right \}.
\end{equation*} 
The expected or observed Fisher information is required to obtain the interval estimations and hypothesis tests on the parameters model. Under standard regularity conditions, $\widehat{\bm{\theta}} - \bm{\theta}$ is asymptotically distributed as a multivariate normal distribution with mean $\bm{0}$ and variance and covariance matrix equal to the inverse of the Fisher information matrix observed, $\bm{H}^{-1}(\bm{\theta})$, where 
\begin{equation}
    \bm{H}(\bm{\theta}) = \begin{bmatrix} 
    \bm{H}_{\beta \beta}(\bm{\theta}) & \bm{H}_{\beta \phi}(\bm{\theta}) \\ 
    \bm{H}_{\phi \beta}(\bm{\theta}) & H_{\phi \phi} (\bm{\theta})
    \end{bmatrix},
\end{equation}
with its elements are given by 
$\bm{H}_{\beta \beta}(\bm{\theta})=\partial^2 \ell(\bm{\theta})/\partial \beta \partial \beta^{\top}$,
$\bm{H}_{\beta \phi}(\bm{\theta}) = \bm{H}^{\top}_{\phi \beta}(\bm{\theta}) =\partial^2 \ell(\bm{\theta})/\partial \beta \partial \phi$, and $H_{\phi \phi}(\bm{\theta}) =\partial^2 \ell(\bm{\theta})/\partial \phi^2$. We first compute the elements of the square matrix of order $p$, $\bm{H}_{\beta \beta}(\bm{\theta})$. For $m,r=1,\ldots,p$ we have that
\begin{eqnarray*}
   \frac{\partial^2 \ell(\bm{\theta})}{\partial \beta_m\partial \beta_r} &=&
   \sum_{i=1}^n
   \left(
   \frac{A_i(\bm{\theta})
   \frac{\partial^2 A_i(\bm{\theta})}{\partial \beta_m\partial \beta_r}-
   \frac{\partial   A_i(\bm{\theta})}{\partial \beta_r}
   \frac{\partial   A_i(\bm{\theta})}{\partial \beta_m}}{A_i(\bm{\theta})^2}
   \right) \\
   &=&   \sum_{i=1}^n
   \left(
   \frac{1}{A_i(\bm{\theta})}
   \frac{\partial^2 A_i(\bm{\theta})}{\partial \beta_m\partial \beta_r}-
    \frac{1}{A_i(\bm{\theta})^2}
   \frac{\partial   A_i(\bm{\theta})}{\partial \beta_r}
   \frac{\partial   A_i(\bm{\theta})}{\partial \beta_m}
   \right)
   .
\label{segundaderivadaell} 
\end{eqnarray*}
where
\begin{equation*}
\label{primeiraderivadaAi}
\frac{\partial A_i(\bm{\theta})}{\partial \beta_m}=
\sum_{\alpha \in \widetilde{K}}\frac{(-1)^{k_{+}+1}\phi}{D_i(\bm{\theta})^{\phi+1}}\frac{\partial }{\partial \beta_m}
D_i(\bm{\theta})\quad, \quad
\frac{\partial D_i(\bm{\theta})}{\partial \beta_m}=
\sum_{j=1}^{m_i}(1-k_j-y_{ij})\mu_{ij}x_{ijm}.
\end{equation*}
and
\begin{equation*}
\label{segundaderivadaAi}
\frac{\partial^2 A_i(\bm{\theta})}{\partial \beta_m\partial \beta_r}=
\sum_{\alpha \in \widetilde{K}}\frac{(-1)^{k_{+}+1}\phi}{D_i(\bm{\theta})^{\phi+2}}
\left(
D_i(\bm{\theta})
\frac{\partial^2 D_i(\bm{\theta})}{\partial \beta_m\partial \beta_r}-(\phi+1)\frac{\partial D_i(\bm{\theta})}{\partial \beta_r}\frac{\partial D_i(\bm{\theta})}{\partial \beta_m}\right),
\end{equation*}
where 
$$\frac{\partial^2 D_i(\bm{\theta})}{\partial \beta_m\beta_r}=\sum_{j=1}^{m_i}(1-k_j-y_{ij})\mu_{ij}x_{ijr}x_{ijm}.$$
We notice that by setting $r=m$ we obtain the main diagonal of the $\bm{H}_{\beta \beta}(\bm{\theta})$ matrix. 
Now we use equation~\eqref{lbeta} to compute 
$$\bm{H}_{\phi \beta}(\bm{\theta})=
\left(
\frac{\partial^2 \ell(\bm{\theta})}{\partial \phi\partial \beta_1},\cdots,\frac{\partial^2 \ell(\bm{\theta})}{\partial \phi\partial \beta_p}
\right).$$
In fact,
$$\frac{\partial^2 \ell(\bm{\theta})}{\partial \phi\partial \beta_r}=
\sum_{i=1}^{n} 
\sum_{\alpha \in \widetilde{K}}
\frac{\partial}{\partial \phi}
\left(
\frac{\phi A_i(\bm{\theta})}{D_i(\bm{\theta})^{\phi+1}}
\right)(-1)^{k^\alpha_{+}+1} 
\sum_{j=1}^{m_i}(1-k_j-y_{ij})
\mu_{ij}x_{ijr},$$
where $\alpha=(k_1^\alpha,\cdots,k_{m_i}^\alpha)$ and $k^\alpha_{+}=\sum_{j=1}^{m_i}k^\alpha_j.$ Finally
$$\frac{\partial}{\partial \phi}\left(\frac{\phi A_i(\bm{\theta})}{D_i(\bm{\theta})^{\phi+1}}\right)=
\sum_{\tau \in \widetilde{K}}(-1)^{k_{+}^\tau}\left[\frac{1-\phi\ln(D_i(\bm{\theta}))}{D_i(\bm{\theta})^{2\phi+1}}-
\frac{\phi+2\phi^2}{D_i(\bm{\theta})^{2\phi+2}}\right].$$
Next, we move on to the second derivative of the log-likelihood function for the parameter $\phi$. More precisely, $H_{\phi \phi}(\bm{\theta})=\partial^2 \ell(\bm{\theta})/\partial \phi^2$ and we obtain that
\begin{equation*} 
\label{derivativetwo}
H_{\phi \phi}(\bm{\theta}) =
\frac{\partial^2 \ell(\bm{\theta})}{\partial \phi^2} = 
\sum_{i=1}^{n}
\left\{\ 
\frac{1}{\phi} + 
\frac{A_i(\bm{\theta})^{\prime\prime}}{A_i(\bm{\theta})}-
\left(
\frac{A_i(\bm{\theta})^{\prime}}{A_i(\bm{\theta})}
\right)^2 
\right \},
\end{equation*}
where
$$A_i(\bm{\theta})^{\prime\prime}=
\frac{\partial^2 A_i(\bm{\theta})}{\partial \phi^2}=
\sum_{\alpha \in \widetilde{K}}
\left \{
(-1)^{k_{+}+1} 
\left(
\frac{2-2\phi \log D_i(\bm{\theta})}{D_i(\bm{\theta})^{\phi+1}}-
\frac{\log^2 D_i(\bm{\theta})}{D_i(\bm{\theta})^{\phi}}-
\frac{\phi(\phi+1)}{D_i(\bm{\theta})^{\phi+2}}
\right )
\right\}$$.

\section{Simulation study}
\label{sec:simula}

In this section, we present the results of the simulation experiments designed to evaluate the asymptotic behavior of the maximum likelihood estimators obtained for the MBerGLGR model. The numerical results are based on $R = 1,000$ Monte Carlo replications for sample sizes $n=30, 60, 90, 120,$ and $150$ observations. Binary random variable vectors, $\bm{y}= (\bm{y}_1,\ldots,\bm{y}_n)^{\top}$ with $\bm{y}_{i}=(y_{i1},\ldots,y_{im_{i}})^{\top}$ for $i=1,\ldots,n$, are generated following the hierarchical structure: $(i) ~y_{ij} | b_i \stackrel{\text{ind}}{\sim} \text{Bernoulli}(u_{ij})$, $(ii) ~\log(-\log(1-u_{ij})) = \beta_0 + \beta_1 x_{ij1} + b_i$, and $(iii) ~b_{i} \stackrel{\text{i.i.d.}}{\sim} \text{GLG}(0, \lambda, \lambda)$ for $j=1,\ldots,m_i$ and $i=1,\ldots,n$.

\begin{figure}[h!]
\centering
\begin{minipage}[b]{0.30\linewidth}
\includegraphics[width=\linewidth]{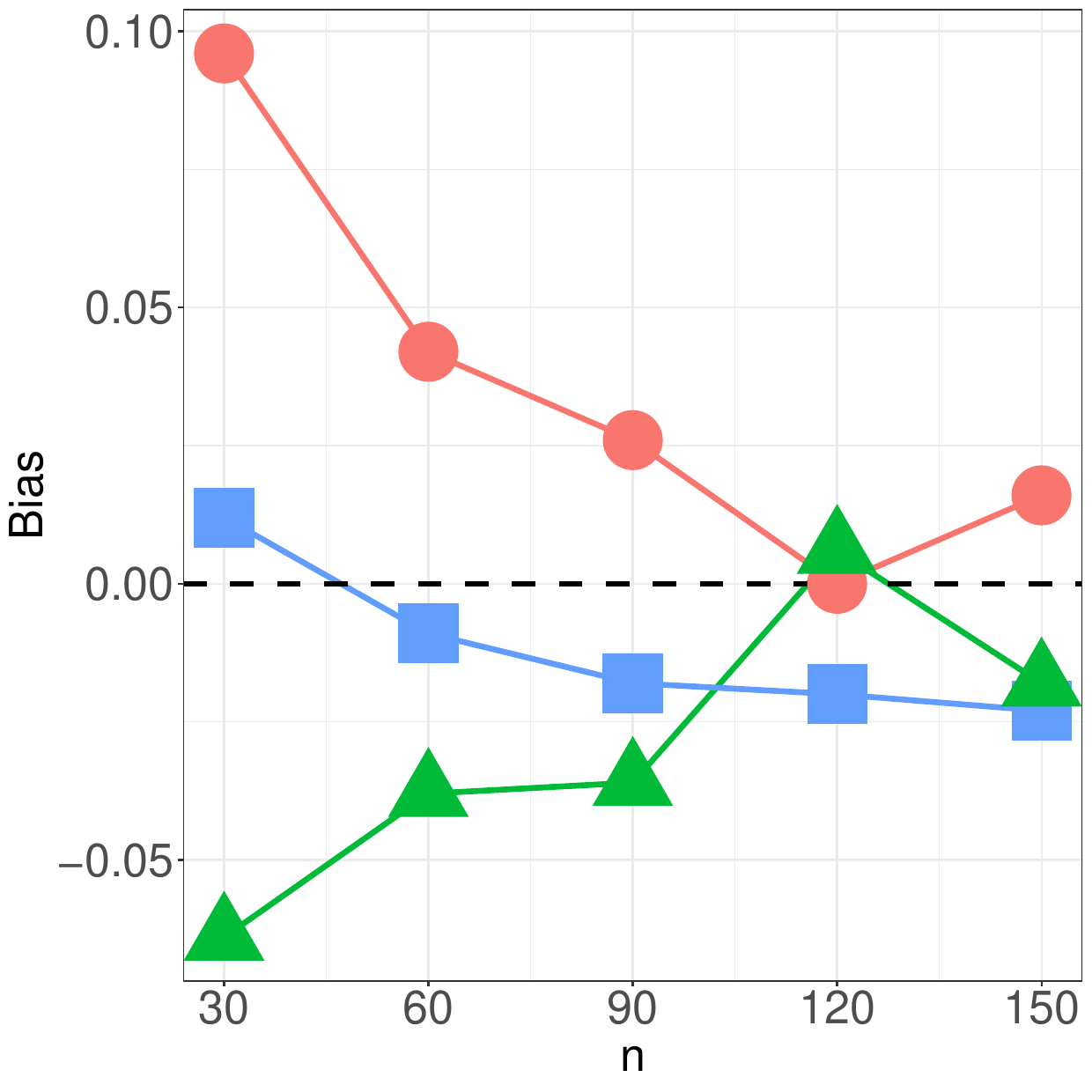}
\end{minipage} 
\hfill
\begin{minipage}[b]{0.30\linewidth}
\includegraphics[width=\linewidth]{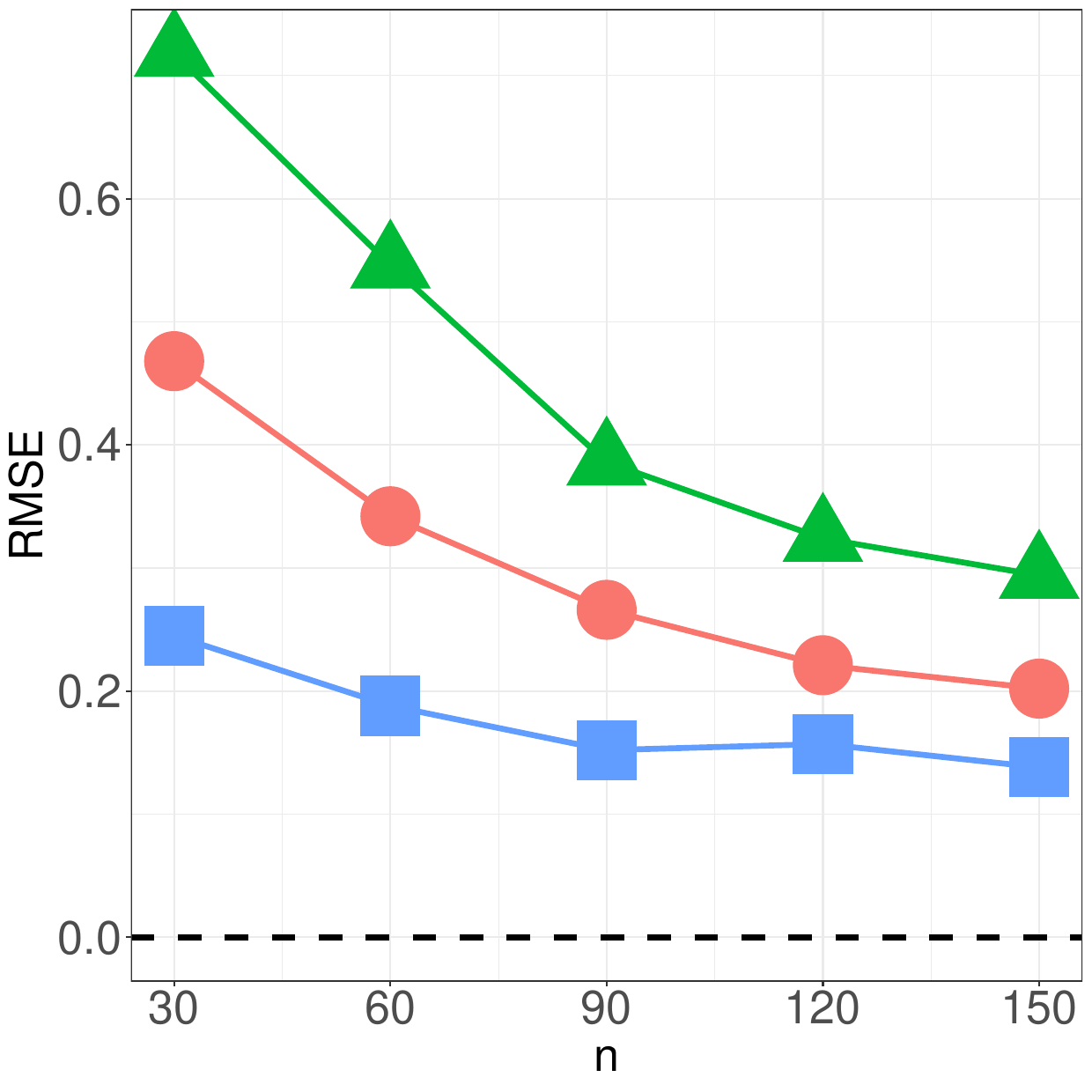}
\end{minipage}
\hfill
\begin{minipage}[b]{0.30\linewidth}
\includegraphics[width=\linewidth]{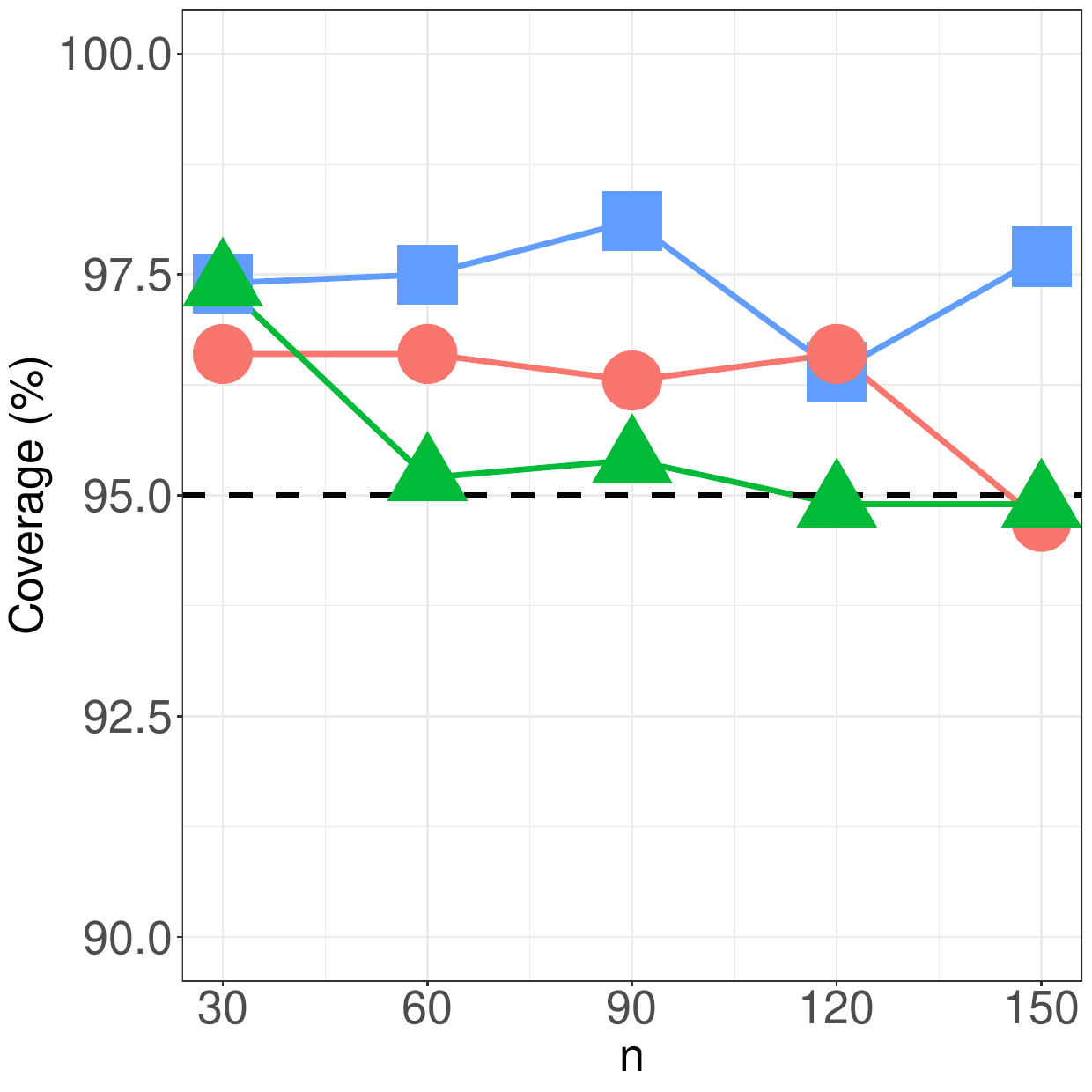}
\end{minipage}\\
\begin{minipage}[b]{0.30\linewidth}
\includegraphics[width=\linewidth]{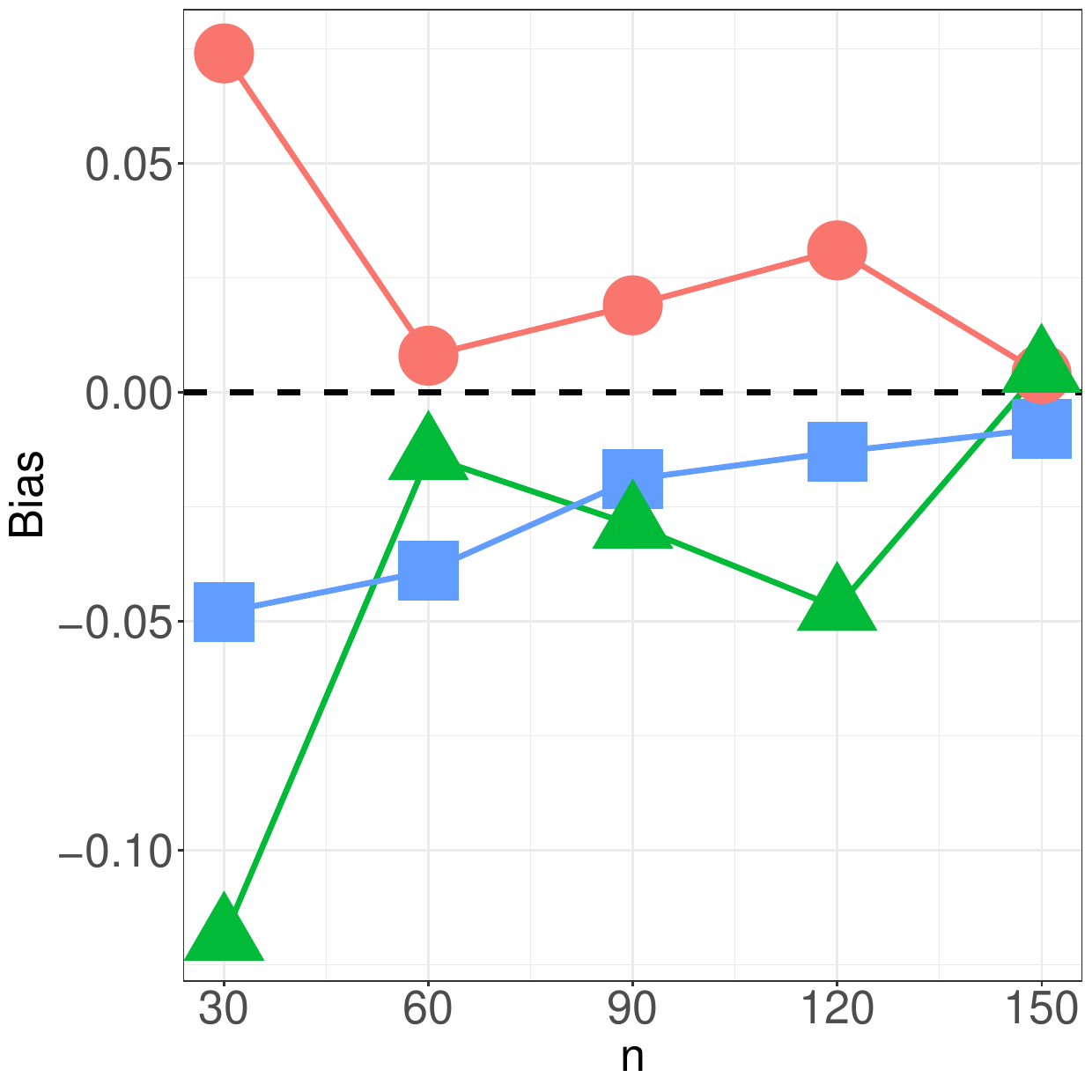}
\end{minipage} 
\hfill
\begin{minipage}[b]{0.30\linewidth}
\includegraphics[width=\linewidth]{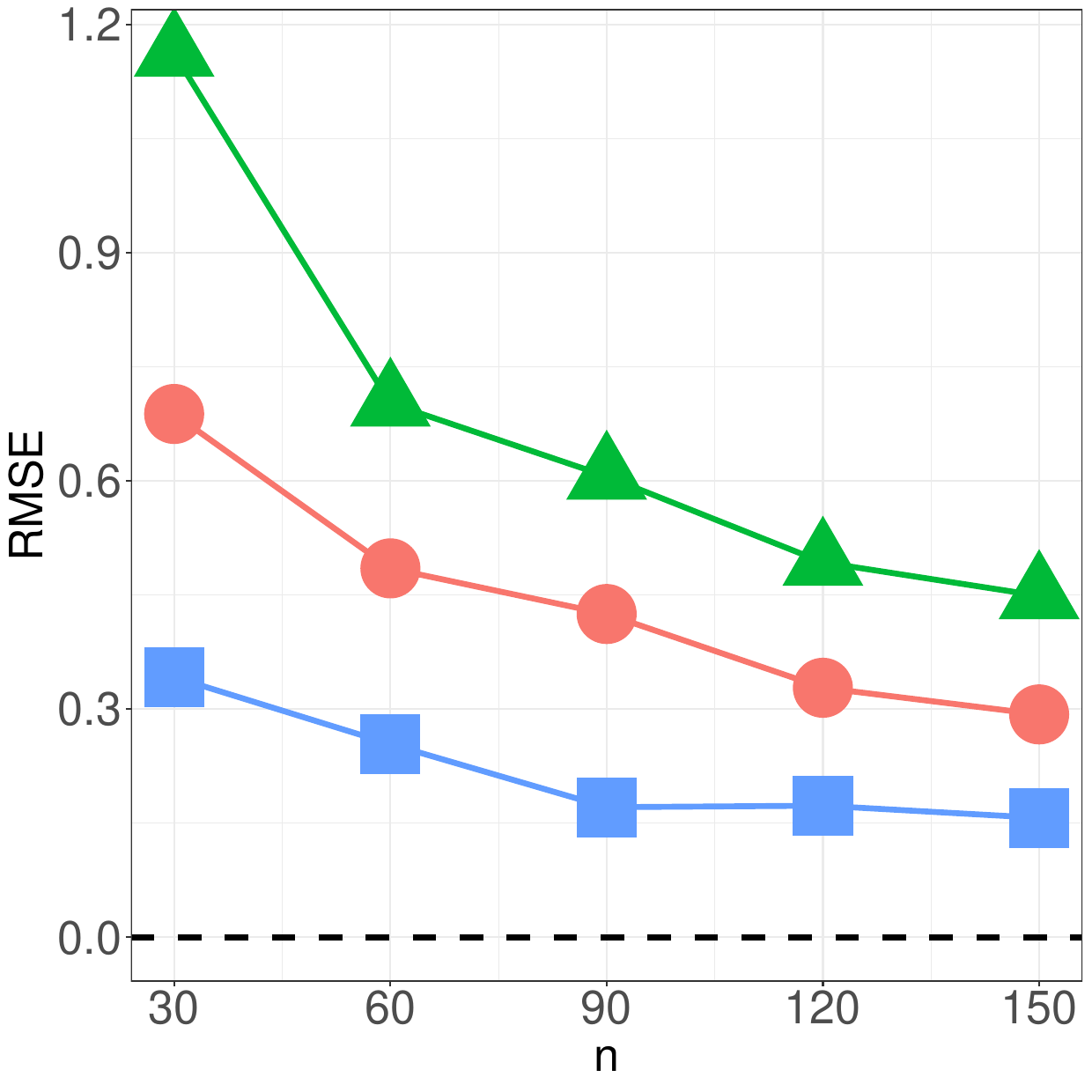}
\end{minipage}
\hfill
\begin{minipage}[b]{0.30\linewidth}
\includegraphics[width=\linewidth]{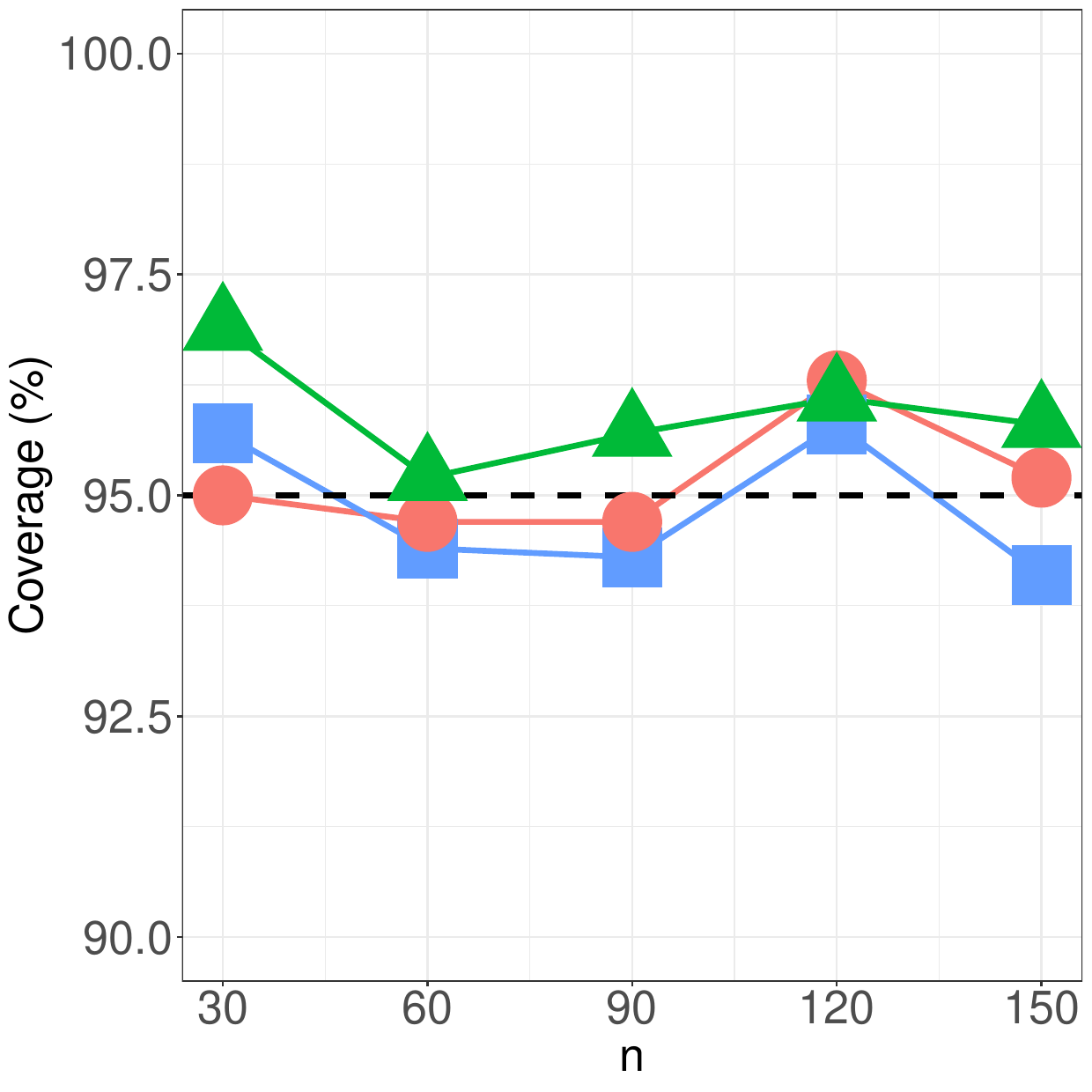}
\end{minipage}\\
\begin{minipage}[b]{0.30\linewidth}
\includegraphics[width=\linewidth]{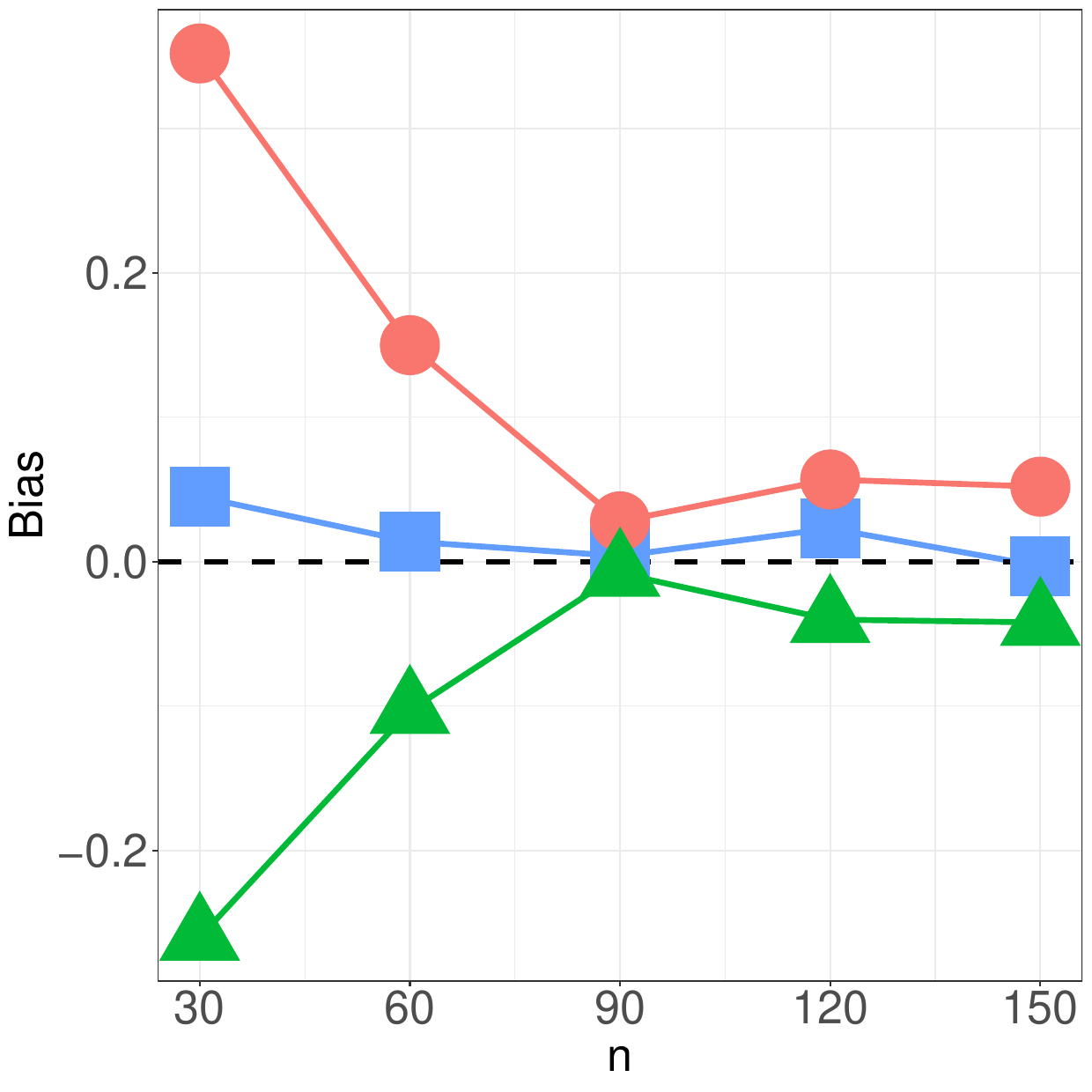}
\end{minipage} 
\hfill
\begin{minipage}[b]{0.30\linewidth}
\includegraphics[width=\linewidth]{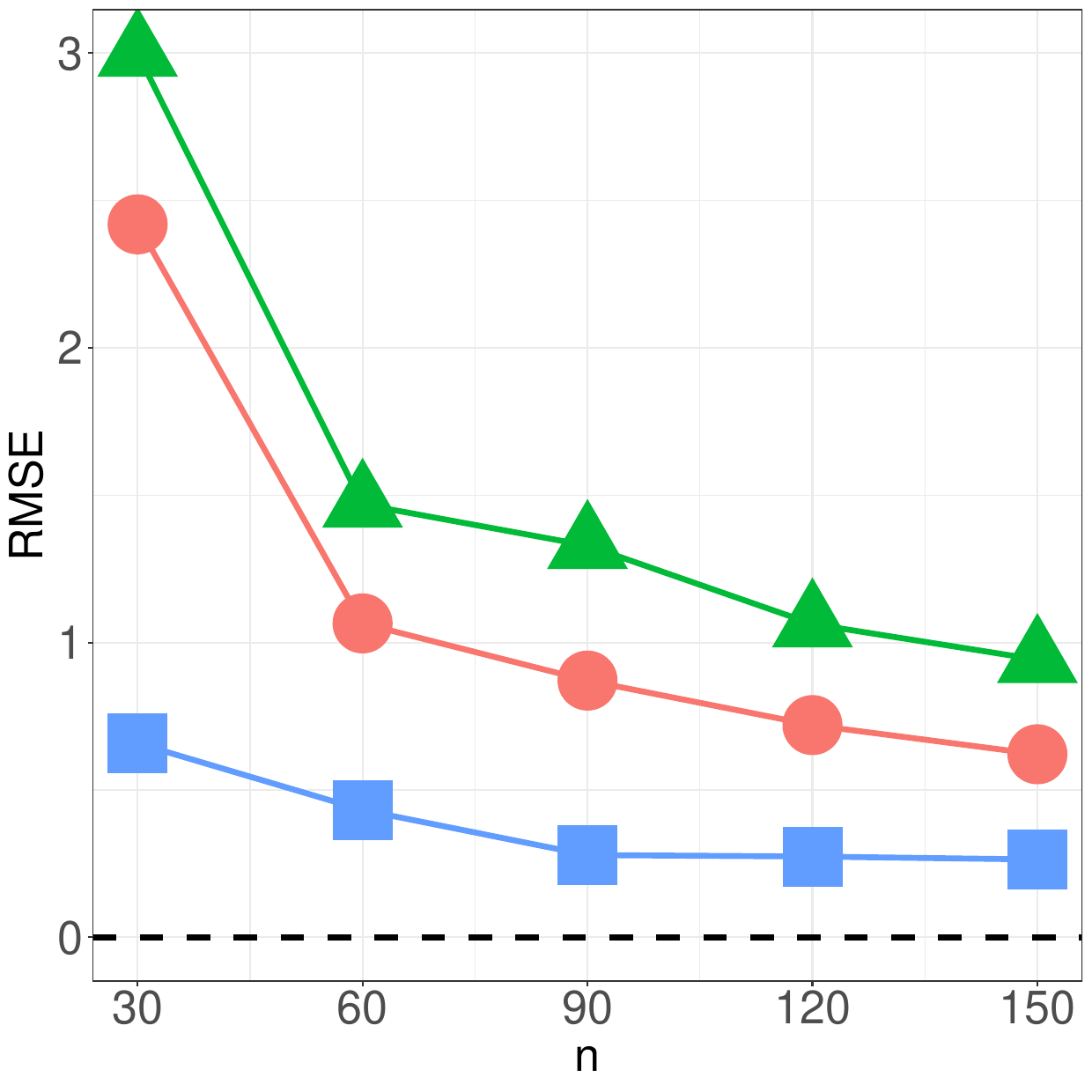}
\end{minipage}
\hfill
\begin{minipage}[b]{0.30\linewidth}
\includegraphics[width=\linewidth]{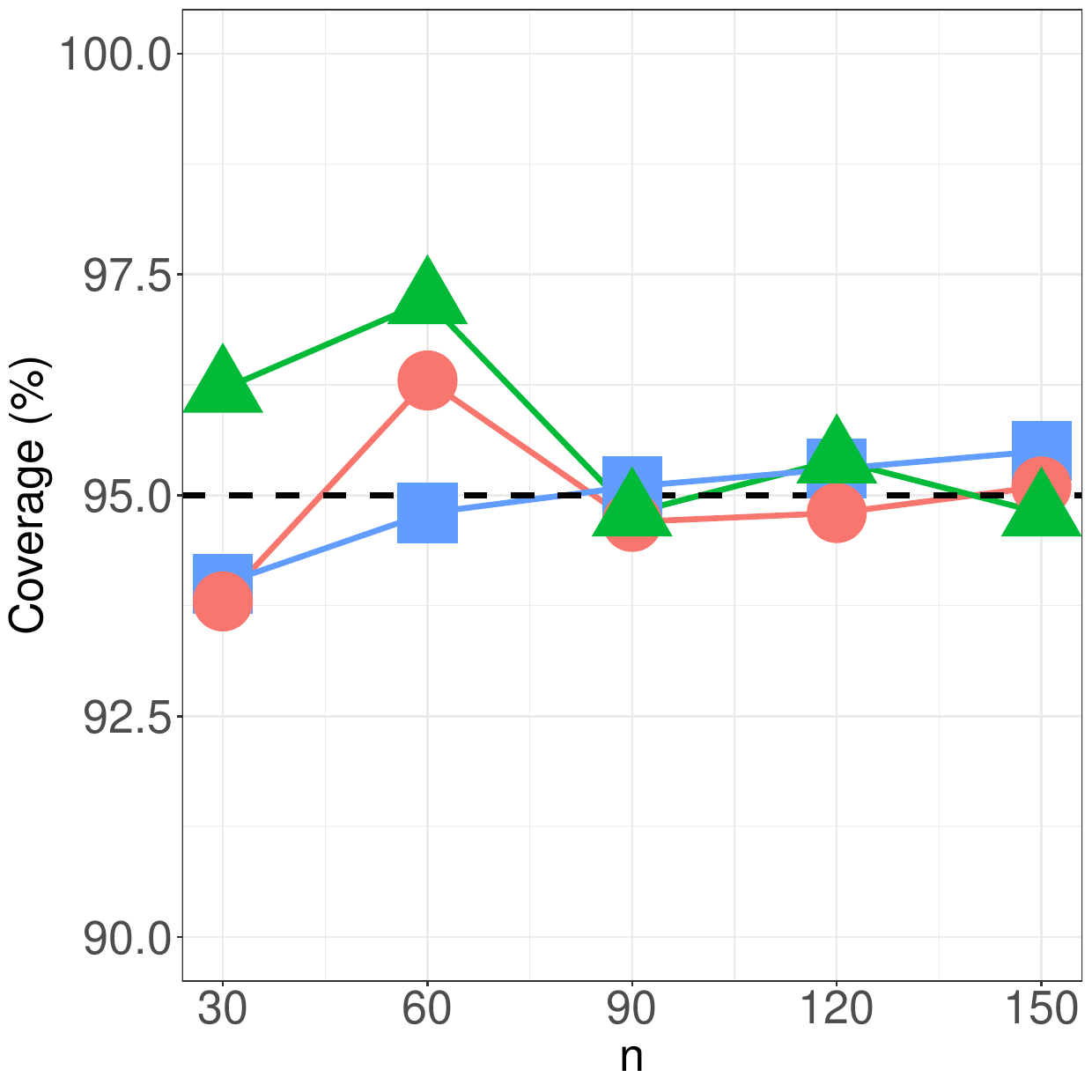}
\end{minipage}
\caption{Bias (left), RMSE (center) and coverage of $95\%$ confidence intervals ($\%$) (right) versus the sample size for estimates of $\lambda$~(square, blue), $\beta_0$~(circle, red) and $\beta_1$~(triangle, green), with $\lambda=$0.5, 1.0 and 3.0(rows) and $x_{ij1} \sim U(0,1)$.}
\label{fig1}
\end{figure}

\begin{figure}[h!]
\centering
\begin{minipage}[b]{0.30\linewidth}
\includegraphics[width=\linewidth]{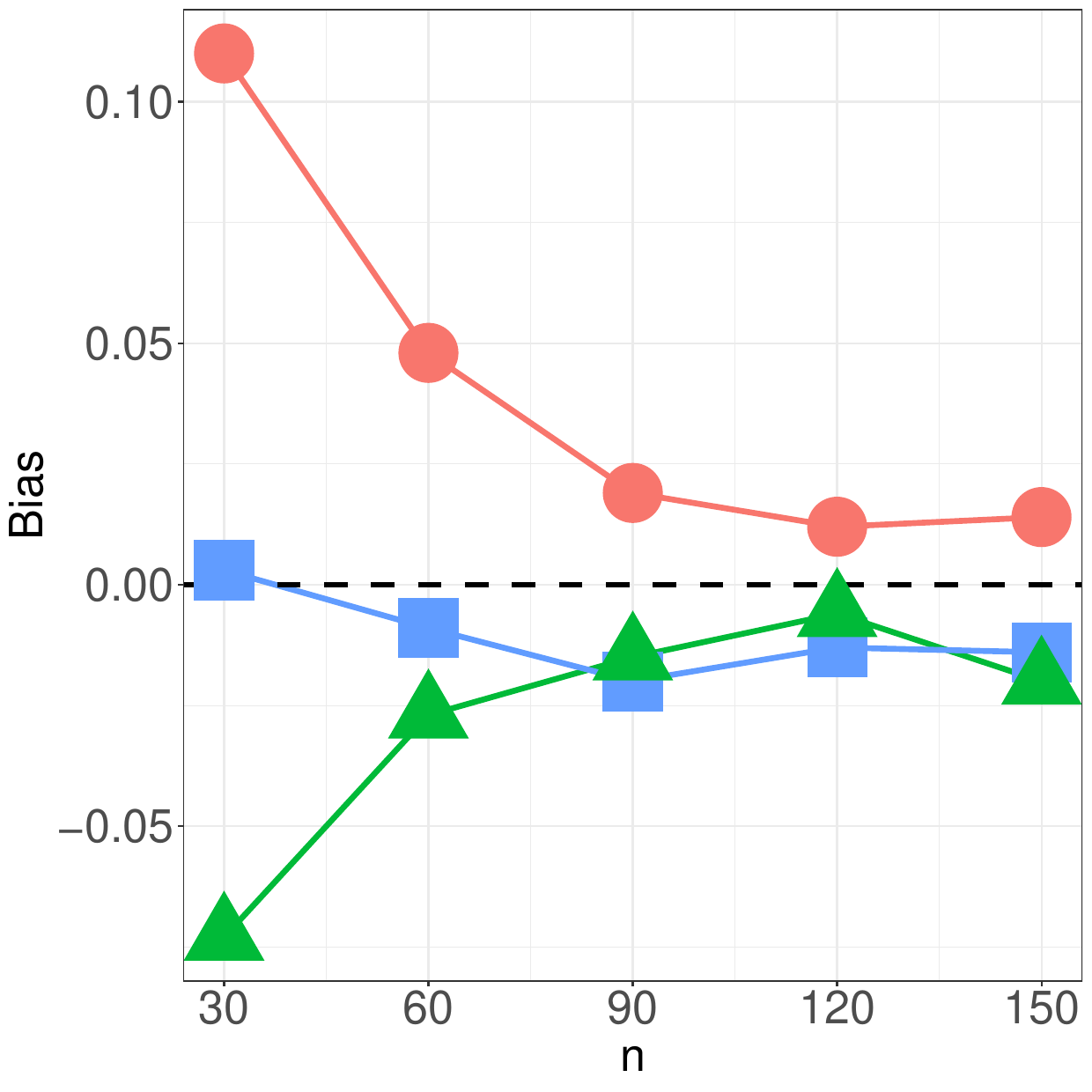}
\end{minipage} 
\hfill
\begin{minipage}[b]{0.30\linewidth}
\includegraphics[width=\linewidth]{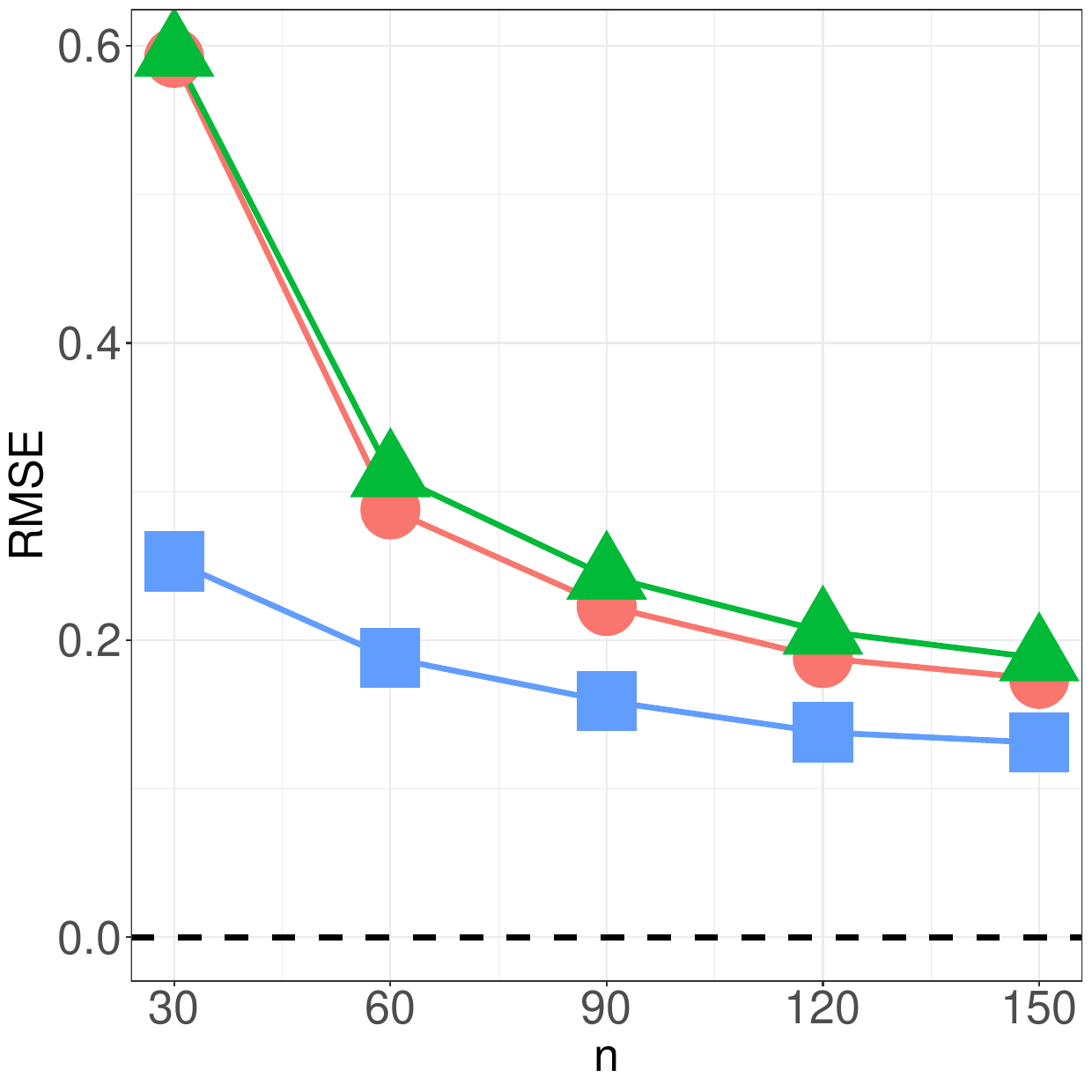}
\end{minipage}
\hfill
\begin{minipage}[b]{0.30\linewidth}
\includegraphics[width=\linewidth]{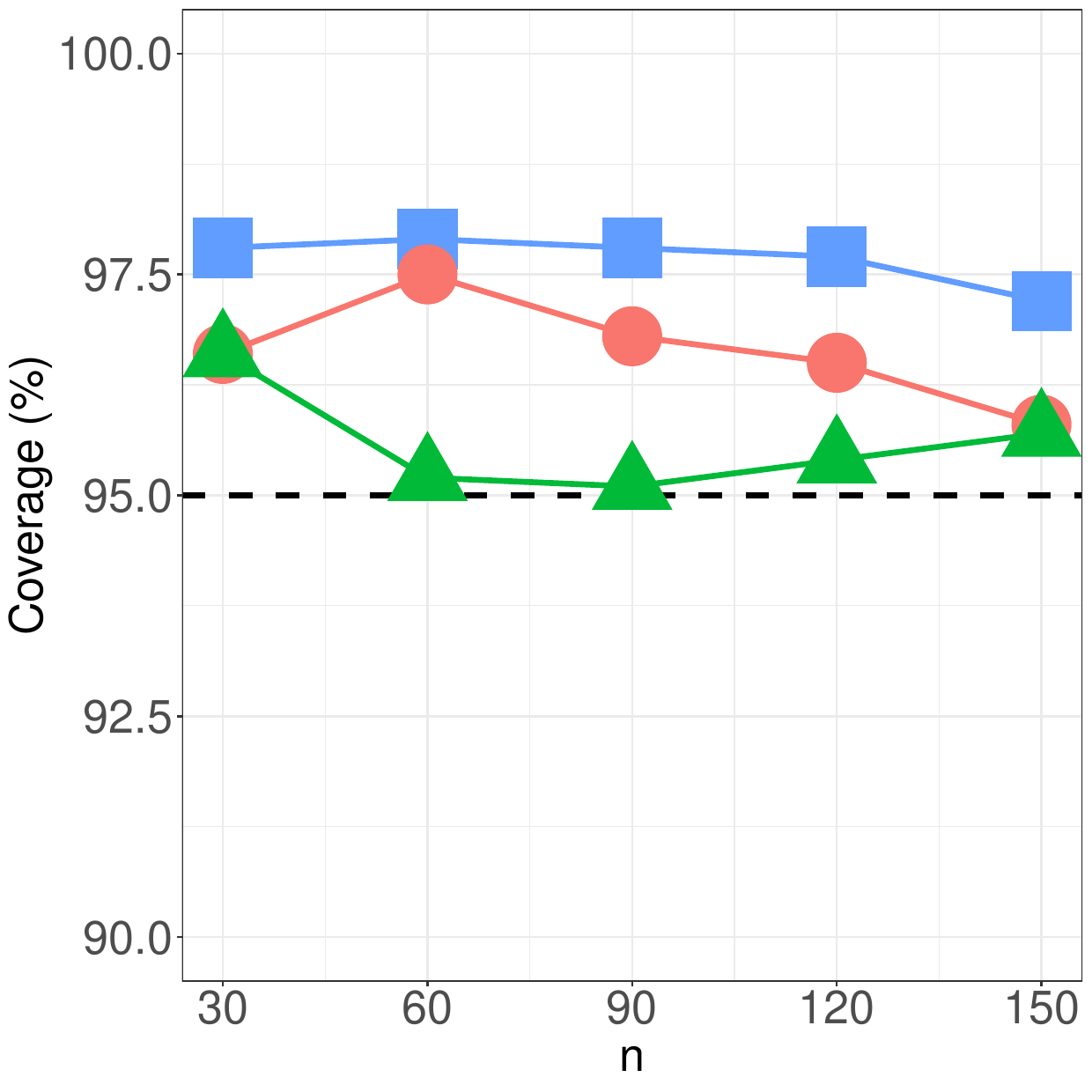}
\end{minipage}\\
\begin{minipage}[b]{0.30\linewidth}
\includegraphics[width=\linewidth]{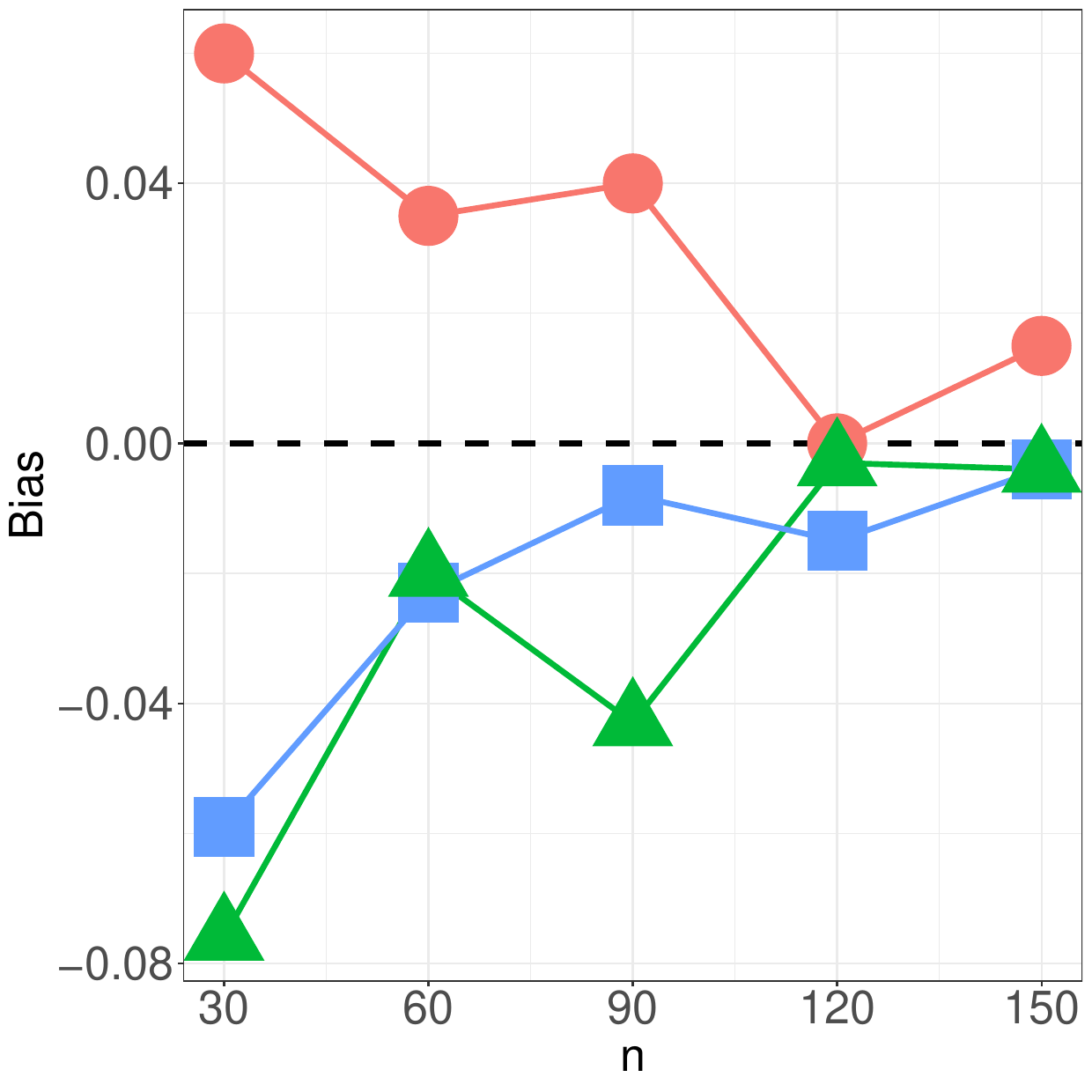}
\end{minipage} 
\hfill
\begin{minipage}[b]{0.30\linewidth}
\includegraphics[width=\linewidth]{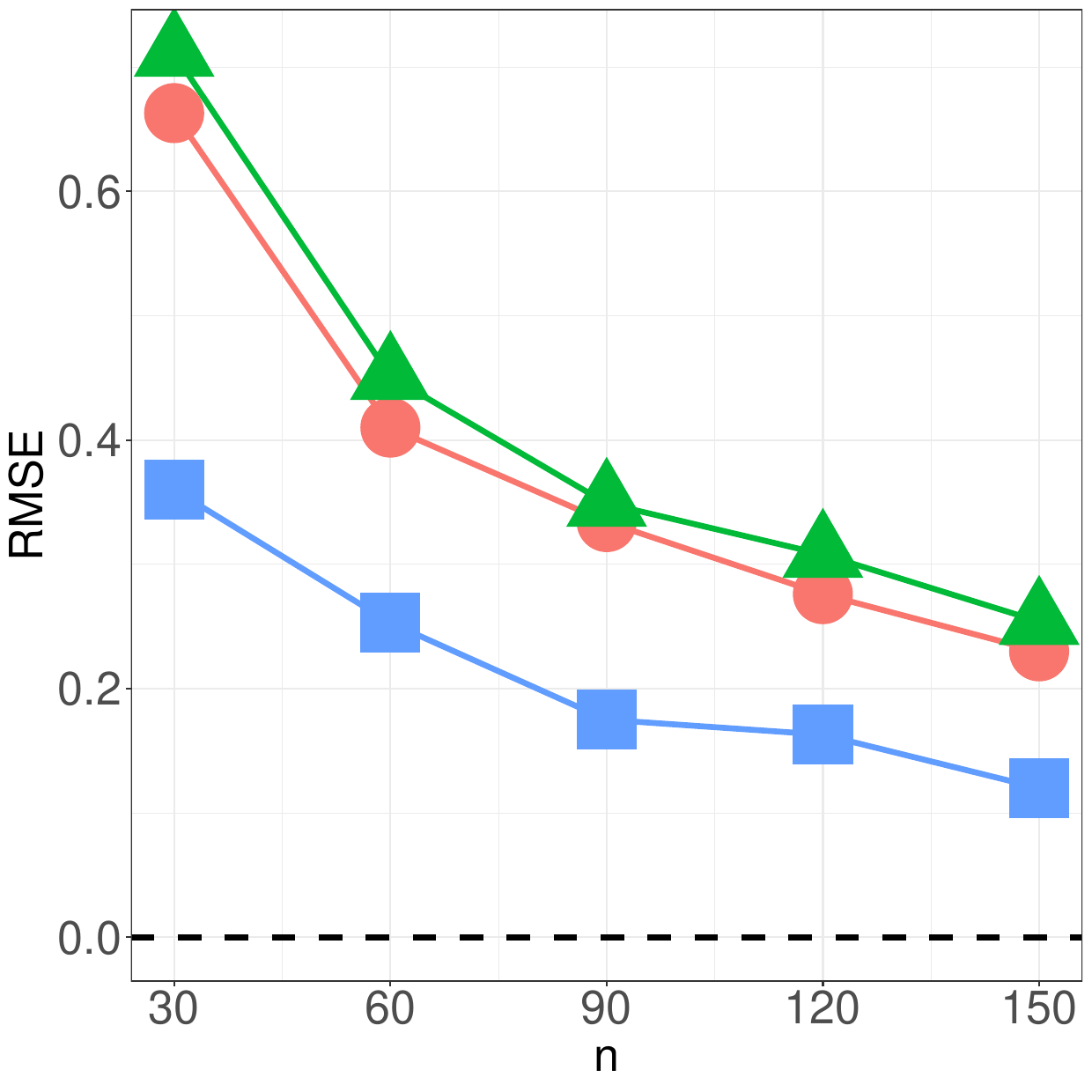}
\end{minipage}
\hfill
\begin{minipage}[b]{0.30\linewidth}
\includegraphics[width=\linewidth]{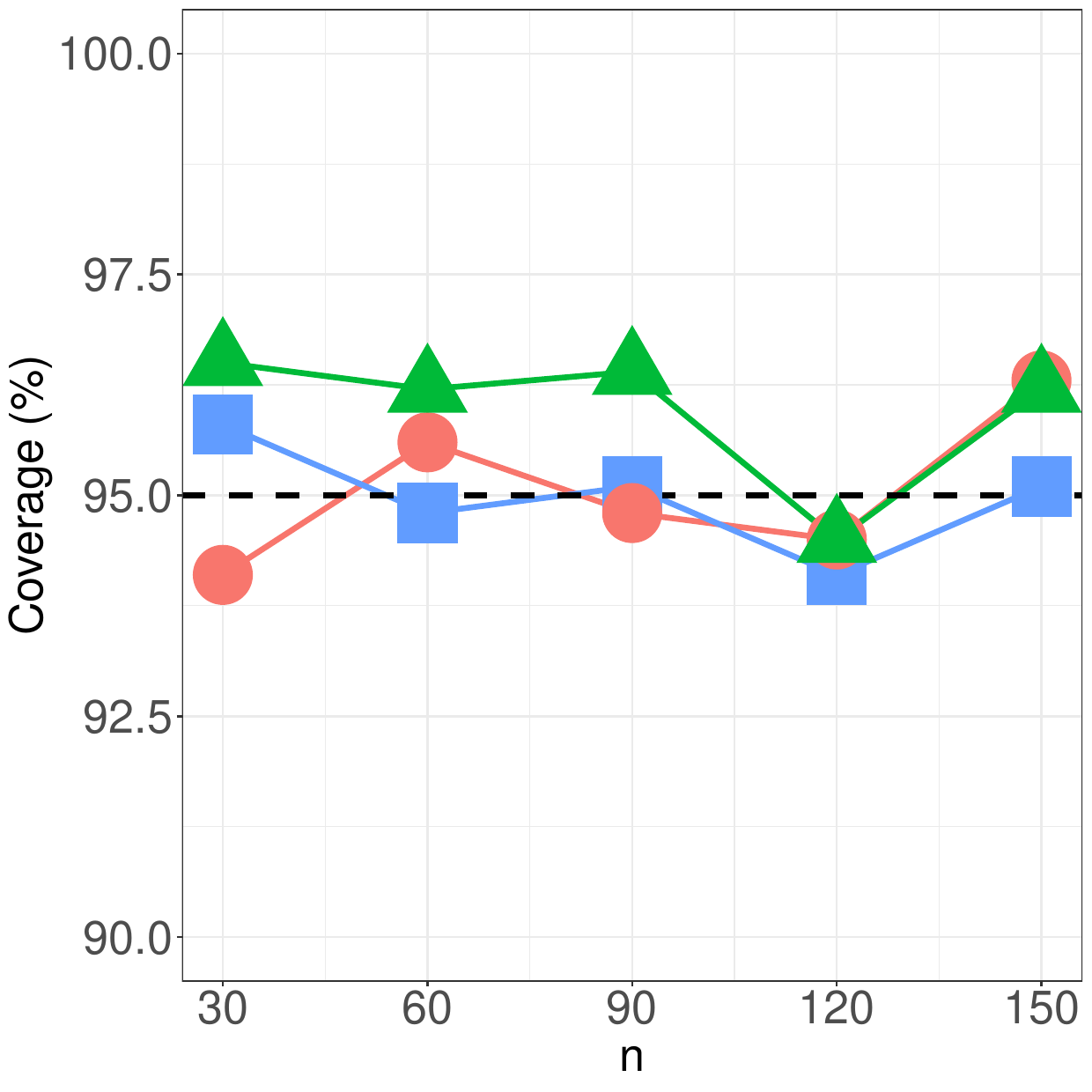}
\end{minipage}\\
\begin{minipage}[b]{0.30\linewidth}
\includegraphics[width=\linewidth]{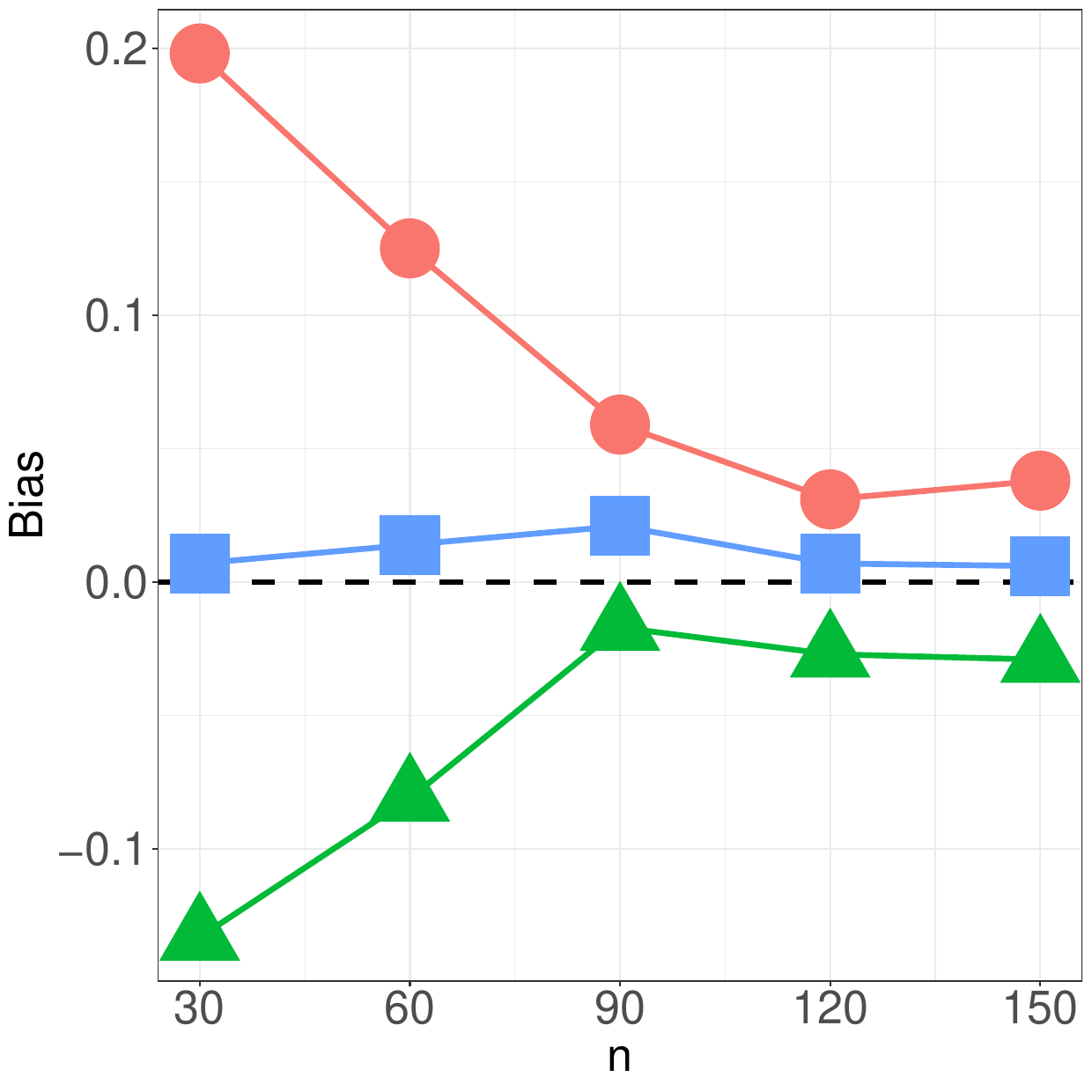}
\end{minipage} 
\hfill
\begin{minipage}[b]{0.30\linewidth}
\includegraphics[width=\linewidth]{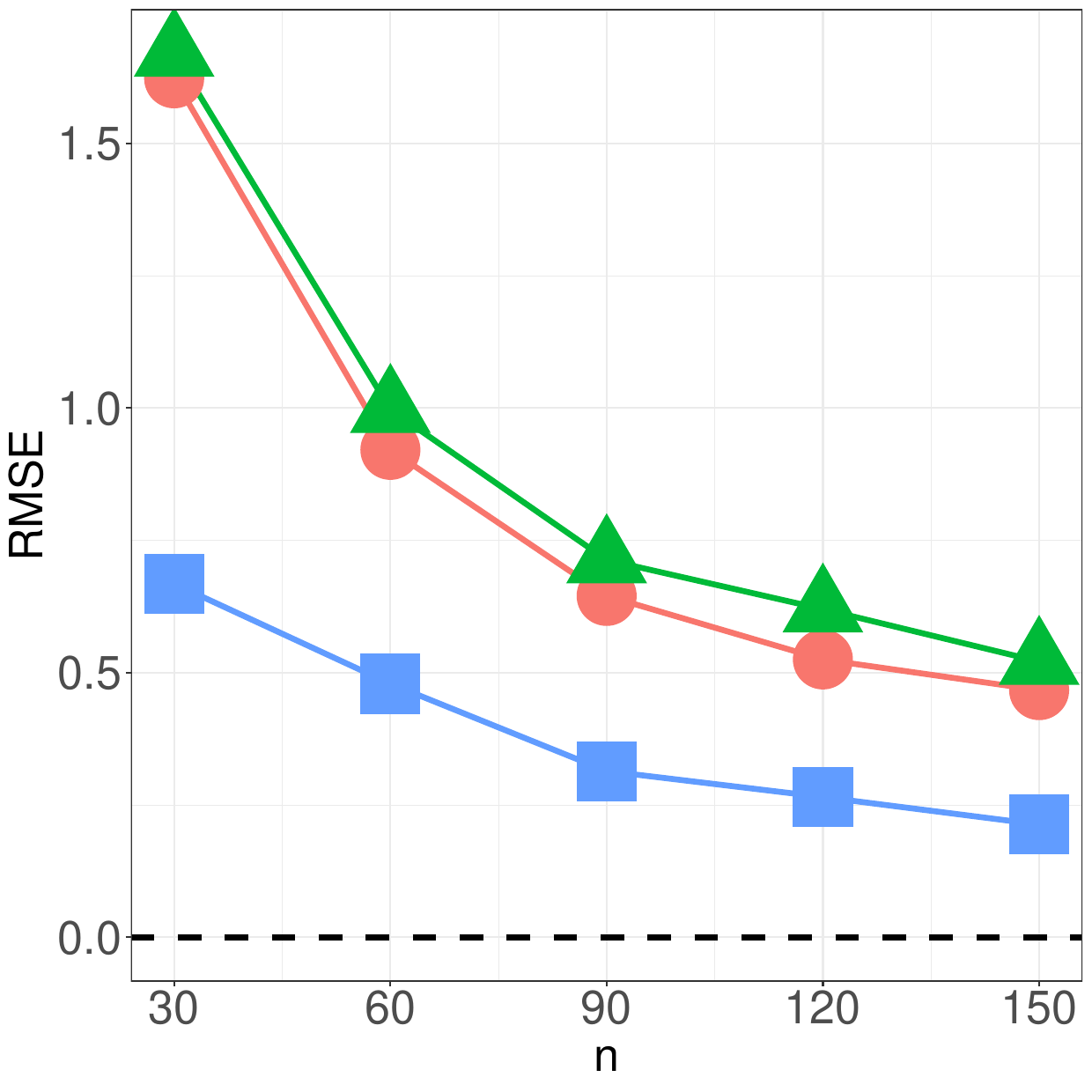}
\end{minipage}
\hfill
\begin{minipage}[b]{0.30\linewidth}
\includegraphics[width=\linewidth]{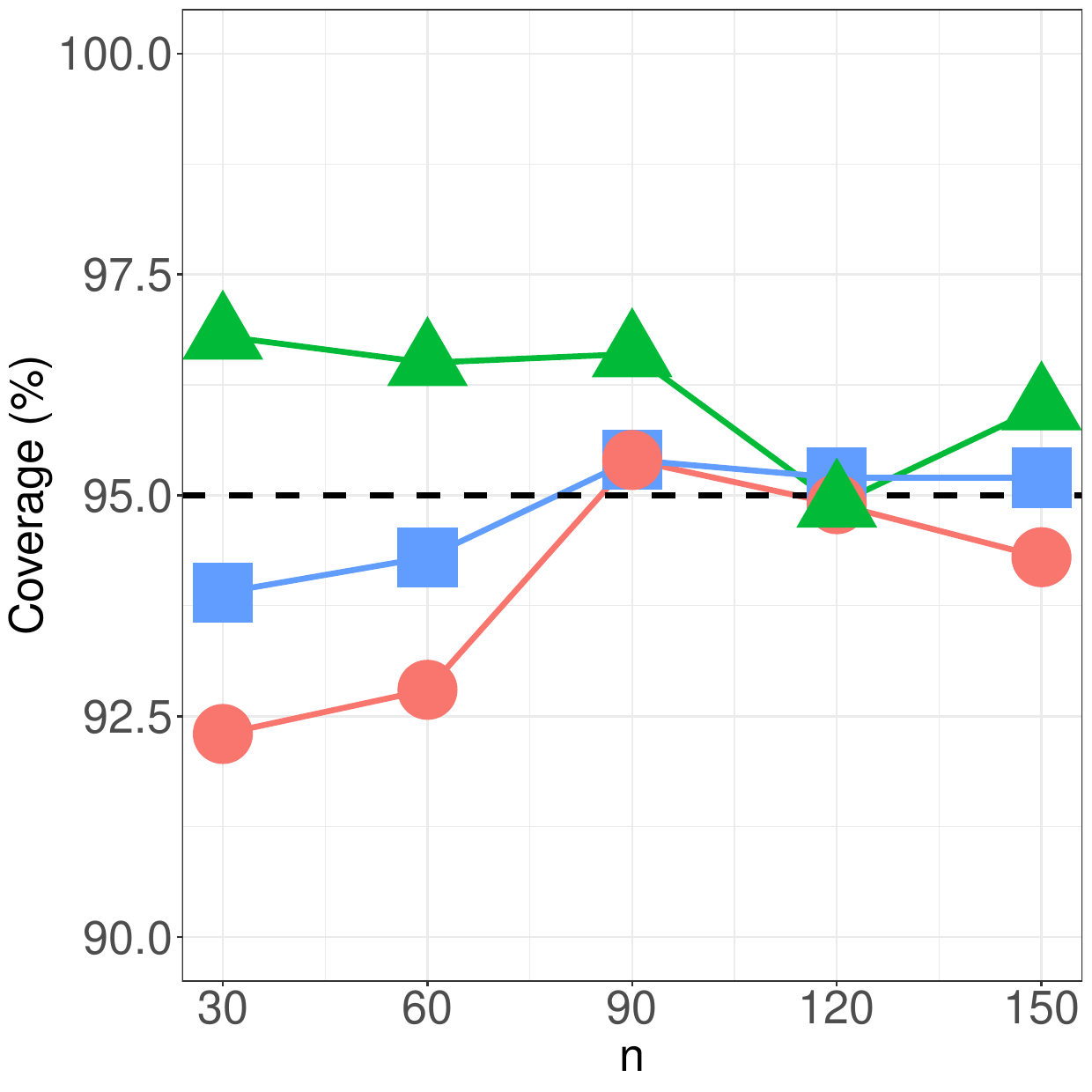}
\end{minipage}
\caption{Bias (left), RMSE (center) and coverage of $95\%$ confidence intervals ($\%$) (right) versus the sample size for estimates of $\lambda$~(square, blue), $\beta_0$~(circle, red) and $\beta_1$~(triangle, green), with $\lambda=$0.5, 1.0 and 3.0(rows) and $x_{ij2}$ as a binary variable.}
\label{fig2}
\end{figure}

Firstly, we consider two scenarios based on the following assumptions: $(i)$ $x_{ij1} \sim U(0,1)$ (scenario 1), and $(ii)$ $x_{ij1}$ as a binary random variable (scenario 2). We assume the parameter set $\bm{\theta}=(\beta_{0},\beta_{1},\lambda)^{\top}=(1.0,-1.0,\lambda)^{\top}$, with three different values for the shape parameter $\lambda=0.5,~1.0$ and $2.0$ in the GLG distribution are considered. The shape parameter values are assumed, considering the number of zeros and ones generated. More precisely, as the $\lambda$ values decrease, the rate of ones generated in $\bm{y}$ increases, and vice versa. The bias, root-mean-squared error (RMSE), and coverage of 95\% confidence intervals were computed. All the simulations were performed using the quasi-Newton {\tt BFGS} nonlinear optimization algorithm implemented in the {\tt optim} function in the software {\tt R} \citep{R-ref}. In Figures \ref{fig1} and \ref{fig2}, we observe that the estimates of the Bias and RMSE  decrease as the sample size $n$ increases, and the coverage probabilities of confidence intervals achieve the level of nominal confidence. We can conclude that the ML estimators have desirable asymptotic properties.

Secondly, three more scenarios are performed to evaluate the asymptotic behavior of the ML estimators from the random intercept Bernoulli-GLG model when the random-effect distribution is misspecified. First, samples are generated from the random intercept Bernoulli-normal model (i.e., $(iii)~b_{i} \stackrel{\text{i.i.d.}}{\sim} \text{N}(0,\lambda)$) and fitted under the Bernoulli-GLG model considering $\lambda=0.5, 10,$ and $2.0$; we are considered as scenario 3 this strategy. Second, samples are generated from the Bernoulli-GLG model and fitted under the Bernoulli-normal model with Probit (i.e., $(ii)~\Phi(u_{ij})=\beta_0 + \beta_1 x_{ij1} + b_i$, scenario 4) and complement log-log (i.e., $(ii)~\log(-\log(1-u_{ij}))=\beta_0 + \beta_1 x_{ij1} + b_i$, scenario 5) link function, respectively, where $x_{ij1} \sim N(0,1)$. The results in Table \ref{tab-simu} reveal that the estimates of the Bias and RMSE increase when the sample assumes higher values. Particularly, scenario 3 indicates that the estimates of the Bias and RMSE related to $\beta_1$ decrease asymptotically, and the $\lambda$ and $\beta_0$ parameters may be considered as biased. This behavior is expected when the $\lambda$ parameter is closer to zero. Further, scenario 5 shows that the Bias and RMSE associated with $\lambda$ are smaller compared with other scenarios, and their regressions coefficient estimates are severely biased. In summary, we conclude that the ML estimators from the random intercept Bernoulli-GLG model are not asymptotically consistent when the random-effect distribution is misspecified.
\begin{table}[h!]
\centering{
\caption{The bias and root-mean-squared error (RMSE) under misspecified of the random effect. }
\label{tab-simu}
\begin{small}
\begin{tabular}{ccrrrrrrrrr}
\cline{3-11}
& & \multicolumn{3}{c}{\multirow{1}{*}{{\rm Scenario 3}}} & \multicolumn{3}{c}{\multirow{1}{*}{{\rm Scenario 4}}} & \multicolumn{3}{c}{\multirow{1}{*}{{\rm Scenario 5}}} \\
\cline{3-11}
& & \multicolumn{3}{c}{\multirow{1}{*}{$b \sim {\rm N}(0,\lambda)$}} & 
     \multicolumn{3}{c}{\multirow{1}{*}{$b \sim { \rm GLG}(0,\lambda,\lambda)$}} & 
     \multicolumn{3}{c}{\multirow{1}{*}{$b \sim {\rm GLG}(0,\lambda,\lambda)$}} \\
& & \multicolumn{3}{c}{\multirow{1}{*}{Bernoulli-GLG}} & 
    \multicolumn{3}{c}{\multirow{1}{*}{Bernoulli-Normal}} & 
    \multicolumn{3}{c}{\multirow{1}{*}{Bernoulli-Normal}} \\ 
& & \multicolumn{3}{c}{\multirow{1}{*}{$g(\mu)=\log(-\log(1-\mu))$}} & 
    \multicolumn{3}{c}{\multirow{1}{*}{$g(\mu)=\Phi(\mu)$}} & 
    \multicolumn{3}{c}{\multirow{1}{*}{$g(\mu)=\log(-\log(1-\mu))$}} \\  
\hline
\multicolumn{11}{c}{\multirow{1}{*}{$\lambda=0.5$}}\\
\hline
Measure & $n$ & $\lambda$ & $\beta_0$ & $\beta_1$ & $\lambda$ & $\beta_0$ & $\beta_1$ & $\lambda$ & $\beta_0$ & $\beta_1$ \\
\hline
\multicolumn{1}{c}{\multirow{5}{*}{Bias}}
& 30     & $-0.026$ &    $0.323$  & $-0.159$       & $0.135$ & $0.000$  & $0.643$        & $0.134$ & $-0.334$  & $0.643$ \\
& 60     & $-0.042$ &   $0.142$  & $-0.019$        & $0.109$ & $-0.036$  & $0.642$       & $0.032$ & $-0.425$  & $0.655$ \\
& 90     & $-0.061$ &    $0.112$  & $-0.006$       & $0.123$ & $-0.037$  & $0.630$       & $0.022$ & $-0.436$  & $0.662$ \\
& 120    & $-0.062$ &    $0.115$  & $-0.016$       & $0.119$ & $-0.055$  & $0.653$       & $0.011$ & $-0.447$  & $0.664$ \\
& 150    & $-0.061$ &    $0.103$  & $0.001$        & $0.122$ & $-0.044$  & $0.628$       & $0.010$ & $-0.455$  & $0.673$ \\
\hline
\multicolumn{1}{c}{\multirow{5}{*}{RMSE}}
& 30     & $0.277$ &    $0.922$  & $0.886$       & $0.553$ & $0.457$  & $0.763$        & $0.754$ & $0.625$  & $0.780$ \\
& 60     & $0.183$ &   $0.319$  & $0.294$        & $0.270$ & $0.256$  & $0.707$        & $0.349$ & $0.516$  & $0.719$ \\
& 90     & $0.161$ &    $0.250$  & $0.228$       & $0.212$ & $0.185$  & $0.668$        & $0.228$ & $0.482$  & $0.697$ \\
& 120    & $0.157$ &    $0.232$  & $0.207$       & $0.188$ & $0.169$  & $0.681$        & $0.141$ & $0.471$  & $0.690$ \\
& 150    & $0.143$ &    $0.203$  & $0.181$       & $0.179$ & $0.141$  & $0.651$        & $0.111$ & $0.470$  & $0.690$ \\
\hline
\multicolumn{11}{c}{\multirow{1}{*}{$\lambda=1.0$}}\\
\hline
\multicolumn{1}{c}{\multirow{5}{*}{Bias}}
& 30     & $-0.277$ &    $0.361$  & $-0.054$       & $0.110$ & $-0.296$  & $0.671$        & $0.220$ & $-0.611$  & $0.606$ \\
& 60     & $-0.221$ &   $0.309$  & $0.001$       & $0.045$ & $-0.367$  & $0.687$        & $0.088$ & $-0.683$  & $0.644$ \\
& 90     & $-0.216$ &    $0.266$  & $0.026$       & $0.042$ & $-0.359$  & $0.676$        & $0.048$ & $-0.740$  & $0.666$ \\
& 120    & $-0.217$ &    $0.252$  & $0.040$       & $0.036$ & $-0.378$  & $0.663$        & $0.014$ & $-0.748$  & $0.674$ \\
& 150    & $-0.206$ &    $0.258$  & $0.039$       & $0.055$ & $-0.385$  & $0.694$        & $0.012$ & $-0.758$  & $0.663$ \\
\hline
\multicolumn{1}{c}{\multirow{5}{*}{RMSE}}
& 30     & $0.451$ &    $0.960$  & $0.867$       & $0.770$ & $0.708$  & $0.905$        & $1.031$ & $0.931$  & $0.979$ \\
& 60     & $0.303$ &   $0.549$  & $0.433$      & $0.288$ & $0.502$  & $0.813$        & $0.612$ & $0.835$  & $0.815$ \\
& 90     & $0.263$ &    $0.414$  & $0.318$       & $0.193$ & $0.427$  & $0.746$        & $0.377$ & $0.821$  & $0.782$ \\
& 120    & $0.251$ &    $0.366$  & $0.282$       & $0.172$ & $0.430$  & $0.715$        & $0.221$ & $0.786$  & $0.740$ \\
& 150    & $0.234$ &    $0.346$  & $0.243$       & $0.158$ & $0.427$  & $0.737$        & $0.151$ & $0.778$  & $0.709$ \\
\hline
\multicolumn{11}{c}{\multirow{1}{*}{$\lambda=2.0$}}\\
\hline
\multicolumn{1}{c}{\multirow{5}{*}{Bias}}
& 30     & $-1.129$ &    $0.860$  & $0.006$       & $0.098$ & $-0.306$  & $0.660$        & $0.265$ & $-0.597$  & $0.644$ \\
& 60     & $-1.383$ &   $0.752$  & $0.041$       & $0.060$ & $-0.340$  & $0.645$        & $0.102$ & $-0.689$  & $0.637$ \\
& 90     & $-1.698$ &    $0.713$  & $0.077$       & $0.039$ & $-0.390$  & $0.694$        & $0.045$ & $-0.723$  & $0.653$ \\
& 120    & $-0.881$ &    $0.685$  & $0.083$       & $0.042$ & $-0.377$  & $0.680$        & $0.019$ & $-0.757$  & $0.681$ \\
& 150    & $-0.712$ &    $0.658$  & $0.109$       & $0.048$ & $-0.371$  & $0.679$        & $0.013$ & $-0.751$  & $0.670$ \\
\hline
\multicolumn{1}{c}{\multirow{5}{*}{RMSE}}
& 30     & $1.518$ &    $1.522$  & $1.195$       & $0.667$ & $0.663$  & $0.874$        & $1.130$ & $0.982$  & $1.012$ \\
& 60     & $1.812$ &   $1.032$  & $0.696$       & $0.365$ & $0.503$  & $0.768$        & $0.621$ & $0.851$  & $0.822$ \\
& 90     & $2.130$ &    $0.896$  & $0.541$       & $0.196$ & $0.458$  & $0.763$        & $0.386$ & $0.793$  & $0.739$ \\
& 120    & $1.110$ &    $0.819$  & $0.456$       & $0.172$ & $0.429$  & $0.735$        & $0.276$ & $0.798$  & $0.747$ \\
& 150    & $0.738$ &    $0.772$  & $0.420$       & $0.157$ & $0.412$  & $0.722$        & $0.163$ & $0.775$  & $0.718$ \\
\hline
\end{tabular}
\end{small}
}
\end{table}
\section{Residual analysis}
\label{sec:residuos}

Residual analysis is used to assess the adequacy of models to data and detect outliers (\citealp{Hilbe2016}). The residual distribution is desirable to follow the standardized normal distribution to check the lack of fitting through graphical methods (\citealp{Pereira20}). \cite{Dunn+Smyth} introduced the randomized quantile residuals for independent continuous or discrete responses, which produces a residual normally distributed. \citealp{Fabio21} suggested the randomized quantile residual for the multivariate negative binomial regression model, once that $y_{i+}= \sum_{i=j}^{m_i}y_{ij},$ where  $\bm{y}_i = (y_{i1},y_{i1}, \dots, y_{im_i})^{\top}$ are independent random response which follows a negative binomial distribution. \citealp{Fabio22} define the randomized quantile residual for the extended inverted Dirichlet (EID) regression model based on conditional variable response. In \cite{Dunn+Smyth} also discussed an extension of randomized quantile residuals for models with dependent responses. Based on these suggestions, \citealp{Fabio22}, we propose the randomized quantile residual for residual analysis of the MBerGLGR model.

Let $\bm{y}_{i}=(y_{i1}, y_{i2}, \ldots, y_{im_{i}})^{\top}$ represent a random vector following a MBerGLG distribution. The randomized quantile residuals \citep{Dunn+Smyth}, which conform to a standard normal distribution, can be employed to evaluate deviations from the MBerGLGR model. If $F(y_{ij}; \mu_{ij},\phi)$ is the cumulative distribution of the probability function defined in (\ref{f10}), then the randomized quantile residuals for $\bm{y}_{ij}$ are given by $r_{q(ij)}=\Phi^{-1}(\nu_{ij})$, where $\Phi(\cdot)$ is the cumulative distribution function of the standard normal and $\nu_{ij}$ is a uniform random variable on the interval $[0,a_{ij}]$ for $y_{ij}=0$ and $[a_{ij},1]$ if $y_{ij}=1$ where $a_{ij}=1-(\phi/(\phi+\mu_{ij}))^{\phi}$. 

\section{Applications}
\label{sec:apli}

\subsection{Arthritis data}
The data set analyzed in this sub-session is discussed in \citet{Fitz95} with the proposal to compare the effectiveness of Auranofin (3 mg. twice daily) in treating patients suffering from rheumatoid arthritis.  During the clinical trial, the patients had five binary self-assessment measurements of arthritis, where self-assessment equals 0 if ``poor'' and 1 if ``good''. Patients had a baseline self-assessment measurement (week 0) and follow-up self-assessment measurements at weeks 1, 5, 9, and 13. Randomization to one of the two treatments, placebo or Auranofin, occurred following the second self-assessment. Figure \ref{fig-plot2}(a) shows that the rate of self-assessment ``good'' is bigger than ``poor''  in both groups.
\begin{figure}[h!]
\centering
\begin{minipage}[b]{0.30\linewidth}
\includegraphics[width=\linewidth]{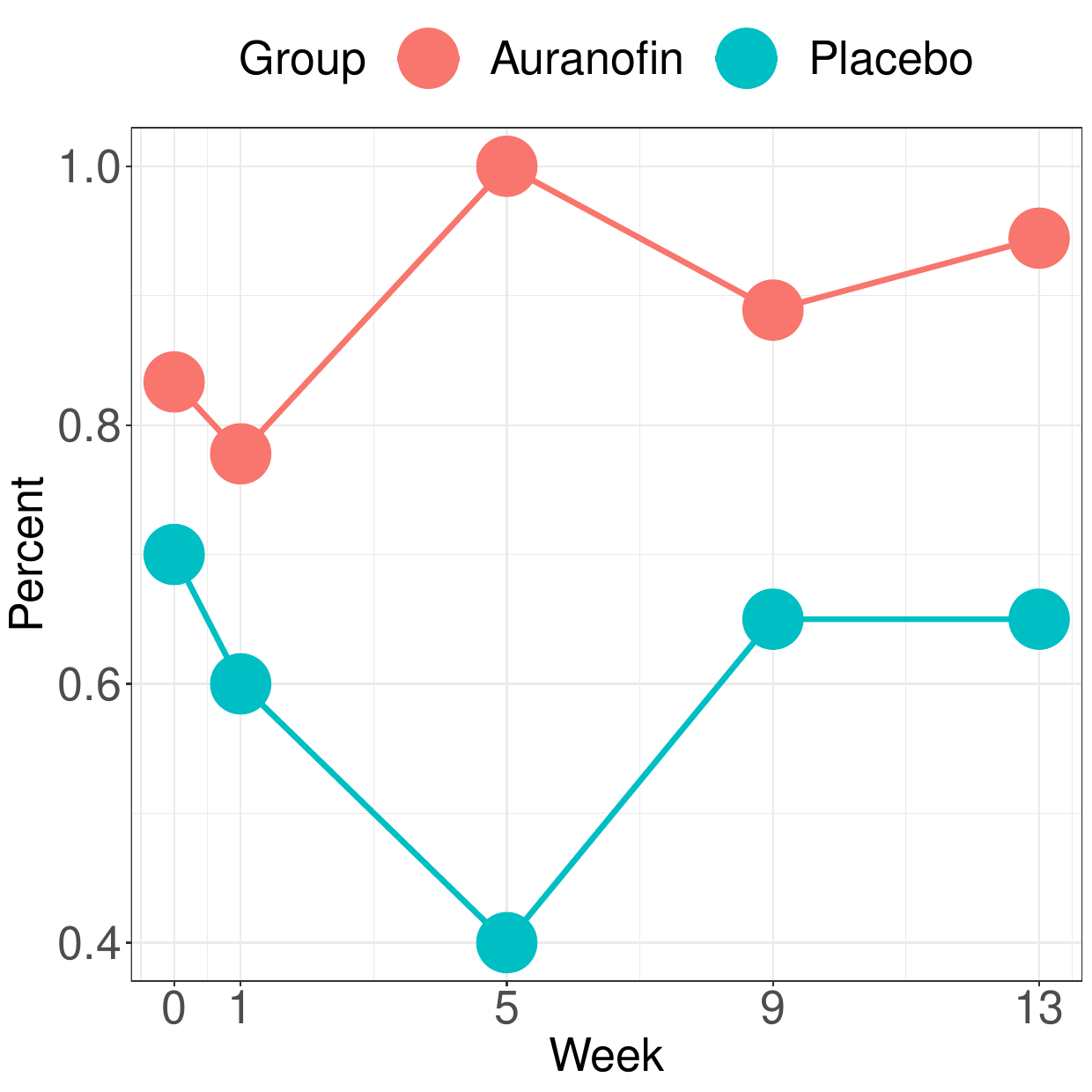}
\begin{center}
    (a)
\end{center}
\end{minipage} 
\begin{minipage}[b]{0.30\linewidth}
\includegraphics[width=\linewidth]{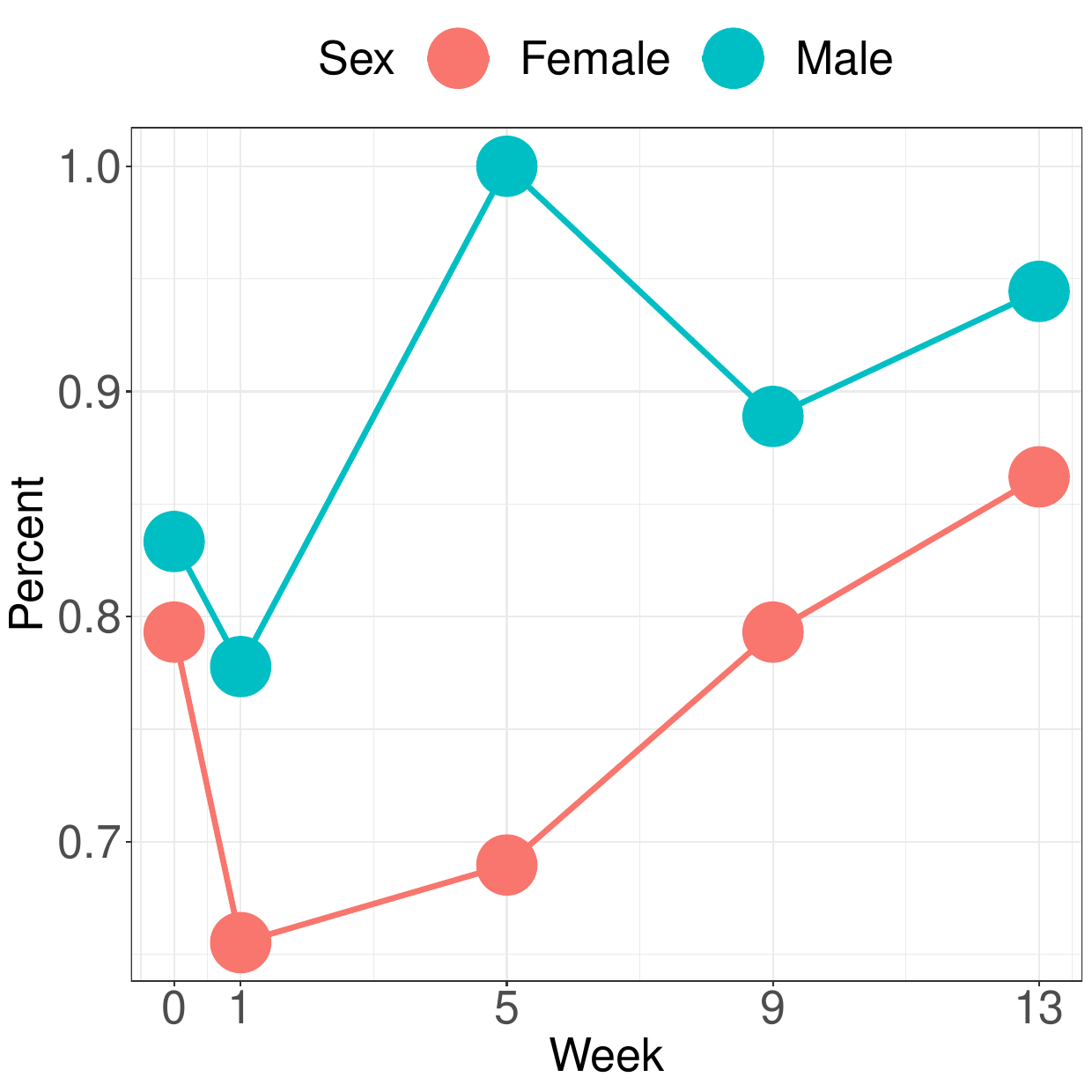}
\begin{center}
    (b)
\end{center}
\end{minipage}
\begin{minipage}[b]{0.30\linewidth}
\includegraphics[width=\linewidth]{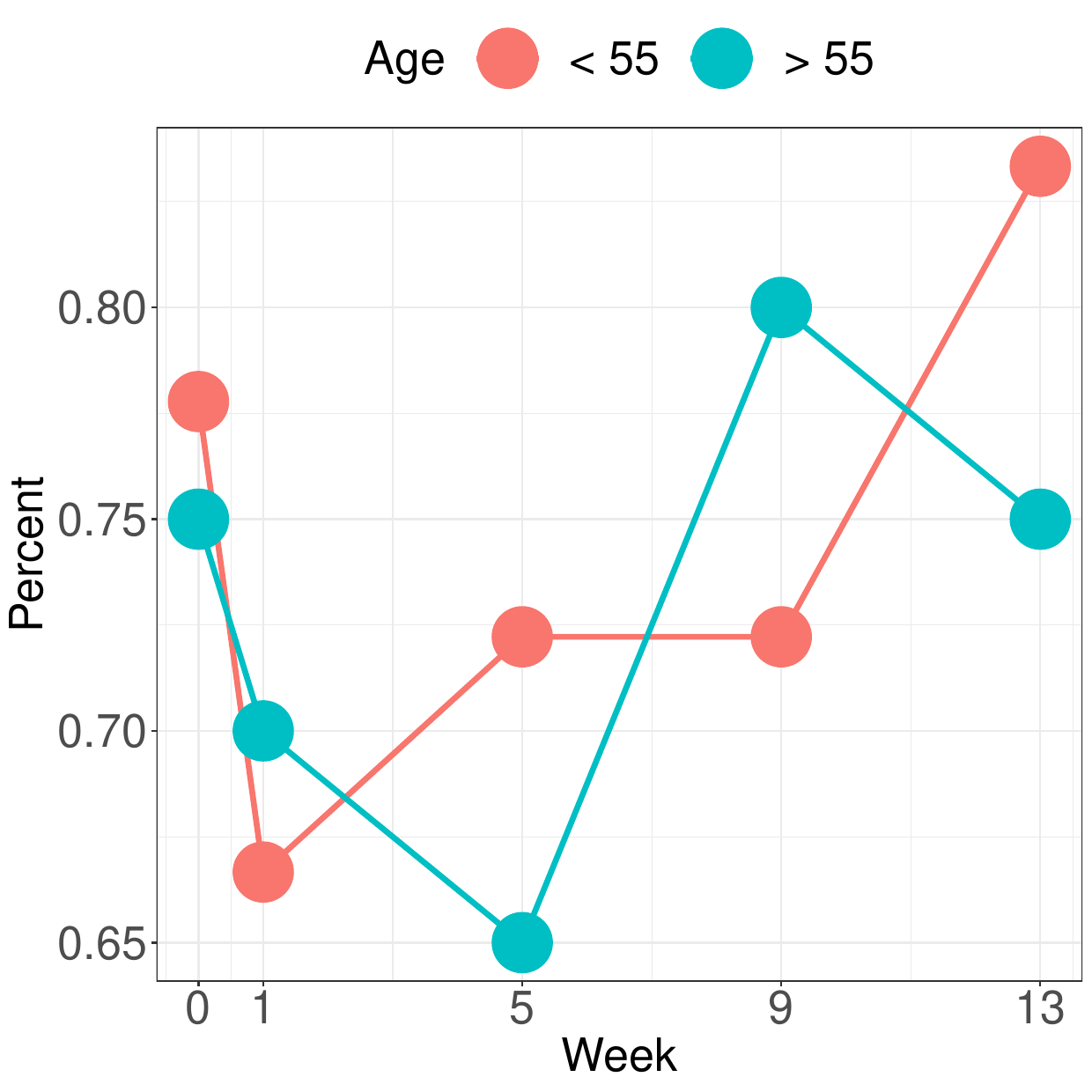}
\begin{center}
    (c)
\end{center}
\end{minipage}
\caption{Percent of patients that self-assessment as ``good'': Treatment(left), Gender (center), and Age (right).}
\label{fig-plot1}
\end{figure}
Figure \ref{fig-plot1}(a) reveals that this behavior occurs over time and that Auranofin can be useful in treating arthritis. Figures~ \ref{fig-plot1}(b) and \ref{fig-plot1}(c) exhibit the rate of self-assessment ``good" associated with the variables gender and age over time, respectively. In general, we observed an atypical behavior in time five. Based on these facts, we proposed the following MBerGLGR model for fitting the Arthritis data.
\begin{eqnarray*}
&(i)&  \bm{y}_i \sim \text{BerGLG}(\bm{\mu}_i, \phi),\\
&(ii)&  \log(-\log(1 - \mu_{ij})) = \beta_0 + \beta_1 ({\rm Sex}_{i1}) + \beta_2 ({\rm Age}_{i2})+ \beta_3 ({\rm Group}_{i3}) + \beta_4 ({\rm Times}_{ij4}), 
\end{eqnarray*}
where $\bm{y}_i = (y_{i1}, \ldots, y_{im_i})^{\top},$ is the vector containing the self--assessment responses of the $i$th subject over time, ${\rm Sex}_{i}$ ( 1=male, 0=female), the ${\rm Age}_{i}$ (0: less than 50 years and 1: otherwise), ${\rm Group}_{i}$ (0=placebo, 1=Auranofin) for the $i$th subject and ${\rm Times}_{ij}$ the time-point where the $j$th measurement was taken on  $i$th subject, for $i=1,\ldots, 51$ and $j=1,\ldots,4.$ Our analyses considered all the pairwise interactions between linear time and the other covariates. However, none were significant, in particular the time-by-treatment interaction. Table \ref{tab1-apli} shows the maximum likelihood estimates, standard errors, and $p$-values of Wald-type tests obtained from the MBerGLGR model, random intercept Bernoulli-normal models by assuming Probit and complement log-log link functions, respectively. Based on AIC criteria, the proposal model fitted better to than data than others, where the effect of the variable Group is more significant. The Odds Ratio $\exp(1.276)$ indicates that the treatment group has more than 3 times the odds of a positive self-assessment compared to the placebo group. The parameter estimate $\lambda$ is associated with the rate of ones present in the variable response and the value $\hat{\phi} = \hat{\lambda}^{-2}=1.506$ as the precision parameter. Figure \ref{fig-plot2}(b) shows that the residuals present an expected behavior with the presence of the atypical subjects. Figure \ref{fig-plot2}(c) confirms the adequacy of the  MBerGLGR model to fit the Arthritis data. Thus, we can conclude that the Auranofin increases the average probability of a positive self-assessment response. The effects of the other covariates do not contribute to the treatment efficacy.
\begin{table}[h!]
\centering{
\caption{Estimates, standard errors for the Arthritis data}
\label{tab1-apli}
\renewcommand{\arraystretch}{1.2}
\begin{tabular}{lrlrlrl}
\hline
 & \multicolumn{2}{c}{\multirow{1}{*}{Bernoulli-GLG model}} & \multicolumn{2}{c}{\multirow{1}{*}{Bernoulli-Normal}} & \multicolumn{2}{c}{\multirow{1}{*}{Bernoulli-Normal}} \\
 & \multicolumn{2}{c}{\multirow{1}{*}{$g(\mu)=\log(-\log(1-\mu))$}} & \multicolumn{2}{c}{\multirow{1}{*}{$g(\mu)=\Phi(\mu)$}} & \multicolumn{2}{c}{\multirow{1}{*}{$g(\mu)=\log(-\log(1-\mu))$}} \\ 
\hline
Parameter & Estimate & Std. error & Estimate & Std. error & Estimate & Std. error \\
\hline
$\lambda$     &    $0.815$  & $-$           & $0.9695$   & $-$        & $1.161$ & $-$ \\
Interc.       &   $-0.180$  & $0.604$       & $-0.1499$  & $0.6256$    & $-0.675$ & $0.662$ \\
Sex           &    $0.654$  & $0.486$       & $0.7133$   & $0.5139$    & $0.899$ & $0.563$ \\
Age           &    $0.116$  & $0.419$       & $0.1210$   & $0.4336$    & $0.095$ & $0.423$ \\
Group         &    $1.276$  & $0.460^{*}$   & $1.5266$   & $0.5005^{*}$    & $1.856$ & $0.519^{*}$ \\
Time          &   $-0.159$  & $0.304$       & $-0.1600$  & $0.3165$    & $-0.154$ & $0.315$ \\
\hline
AIC           & \multicolumn{2}{c}{\multirow{1}{*}{$187.97$}} &  \multicolumn{2}{c}{\multirow{1}{*}{$188.20$}} & \multicolumn{2}{c}{\multirow{1}{*}{$189.60$}}  \\
\hline
\multicolumn{7}{l}{\multirow{1}{*}{$*$: Signif. variable}}
\end{tabular}
}
\end{table}
\begin{figure}[h!]
\centering
\begin{minipage}[b]{0.30\linewidth}
\includegraphics[width=\linewidth]{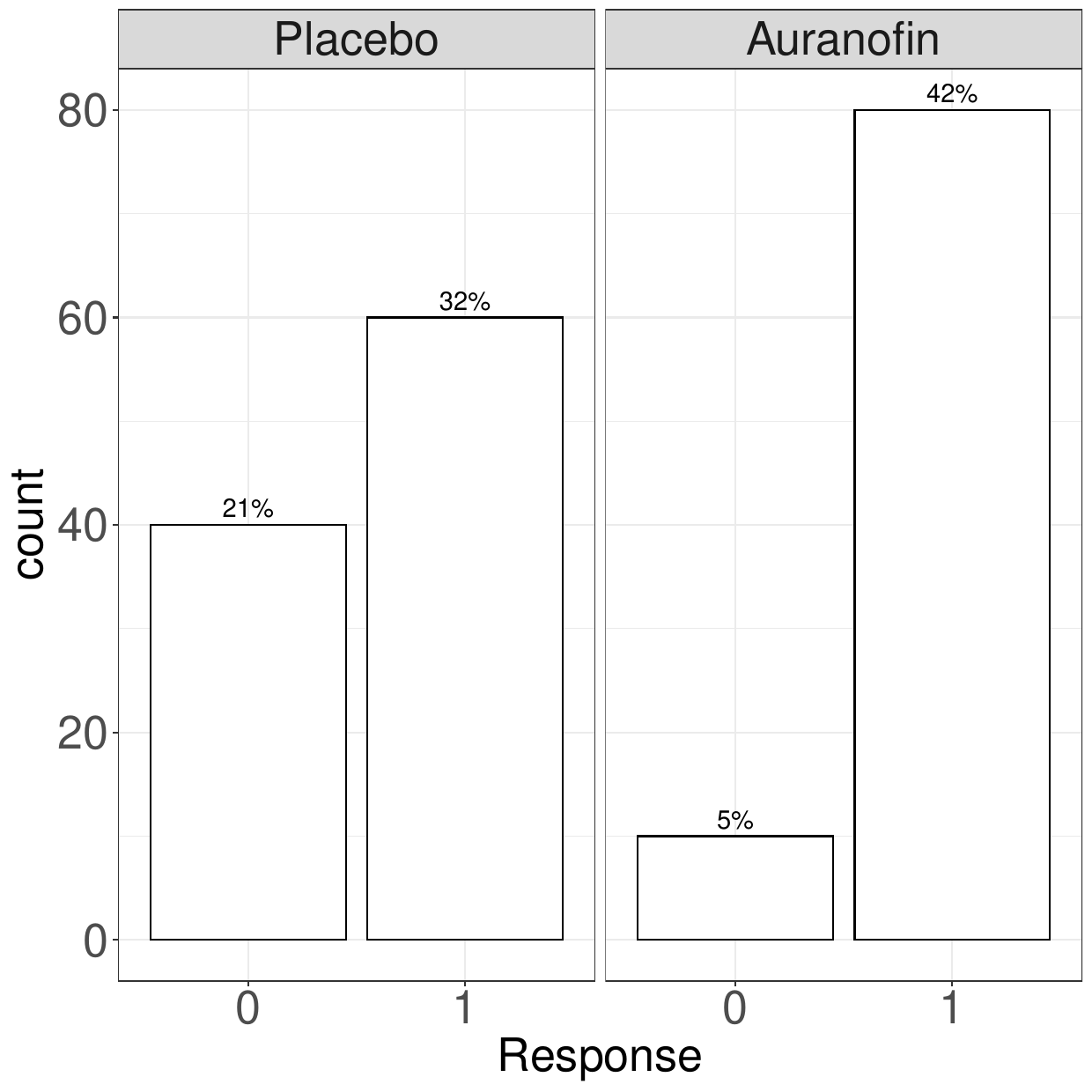}
\begin{center}
    (a)
\end{center}
\end{minipage} 
\begin{minipage}[b]{0.30\linewidth}
\includegraphics[width=\linewidth]{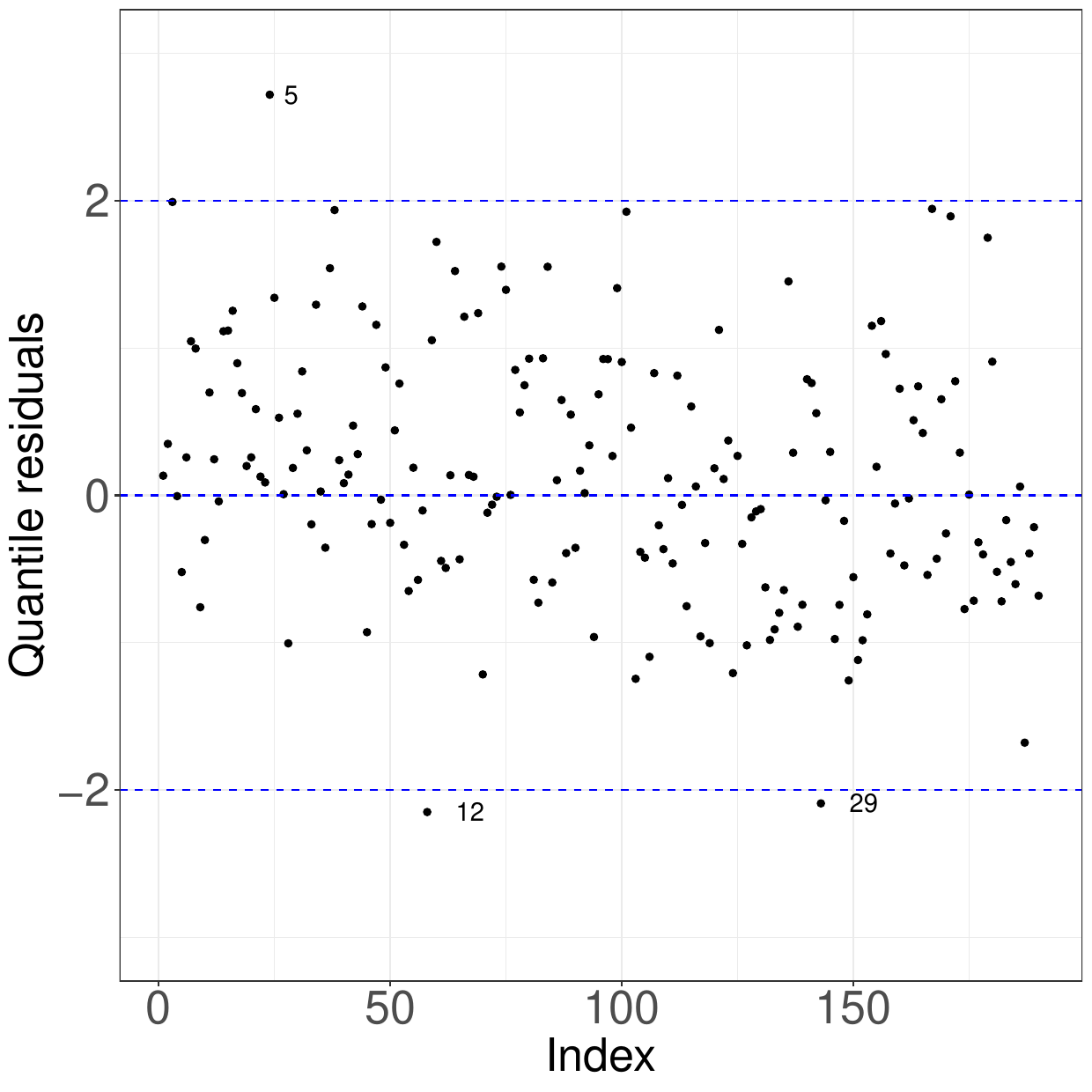}
\begin{center}
    (b)
\end{center}
\end{minipage} 
\begin{minipage}[b]{0.30\linewidth}
\includegraphics[width=\linewidth]{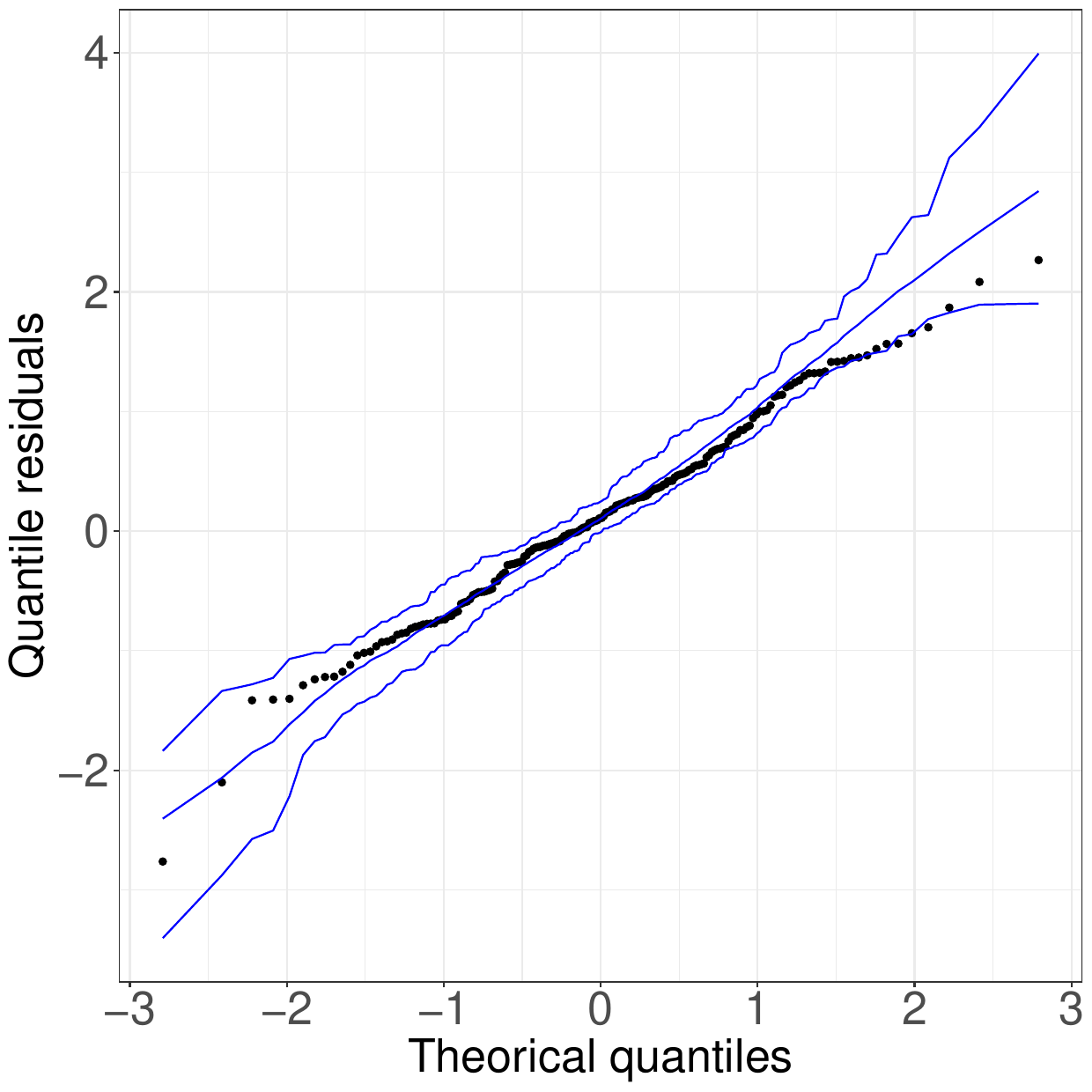}
\begin{center}
    (c)
\end{center}
\end{minipage}
\caption{Bar plot (panel left), Quantile residuals versus index observations (panel center), and simulation envelope (panel right) for the Arthritis data.}
\label{fig-plot2}
\end{figure}

\subsection{Toenail data}

\begin{figure}[h!]
\centering
\begin{minipage}[b]{0.30\linewidth}
\includegraphics[width=\linewidth]{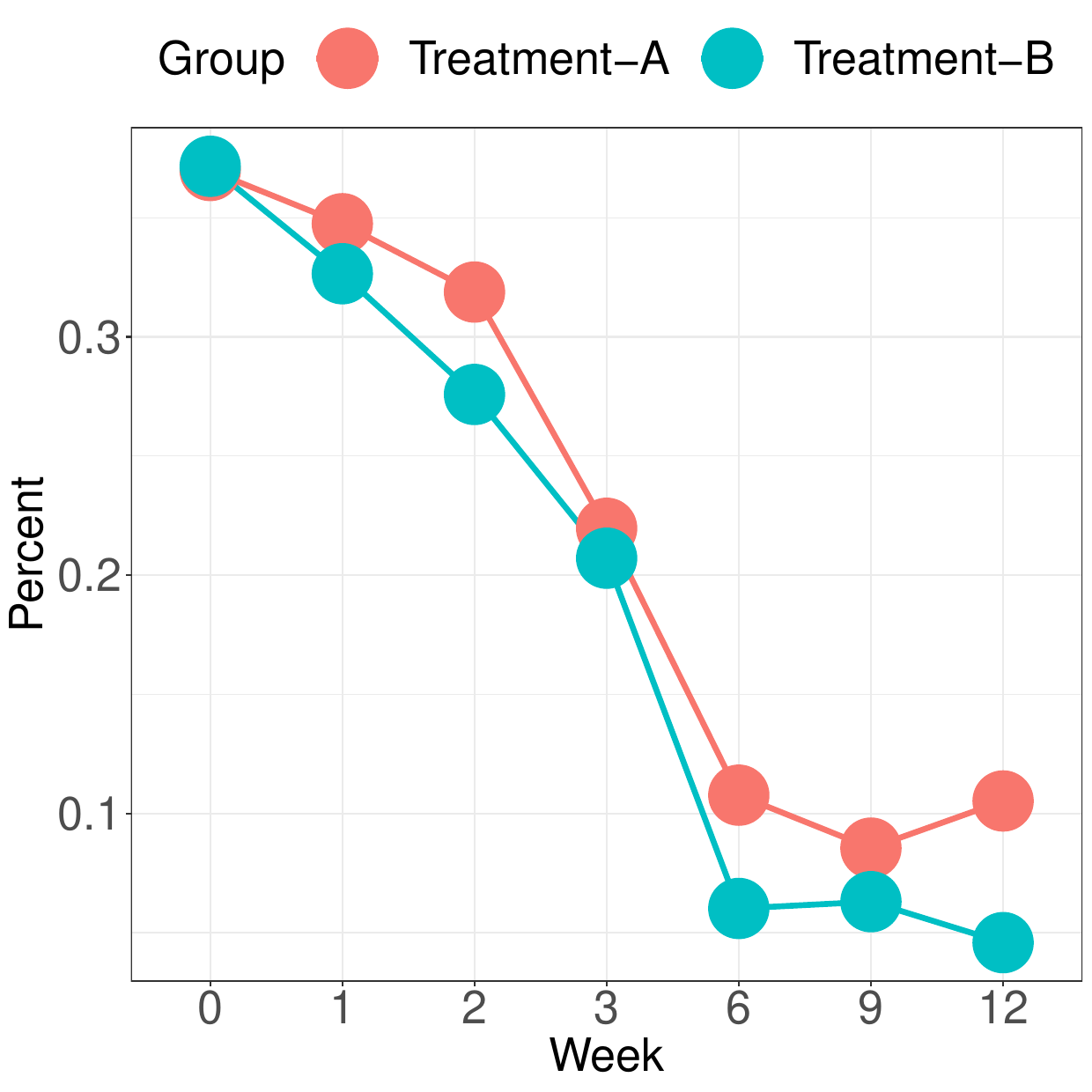}
\end{minipage}
\caption{Percent of patients with responses as the severity of infection, 0 (not severe) and 1 (severe).}
\label{groupToenail}
\end{figure}

\begin{figure}[h!]
\centering
\begin{minipage}[b]{0.30\linewidth}
\includegraphics[width=\linewidth]{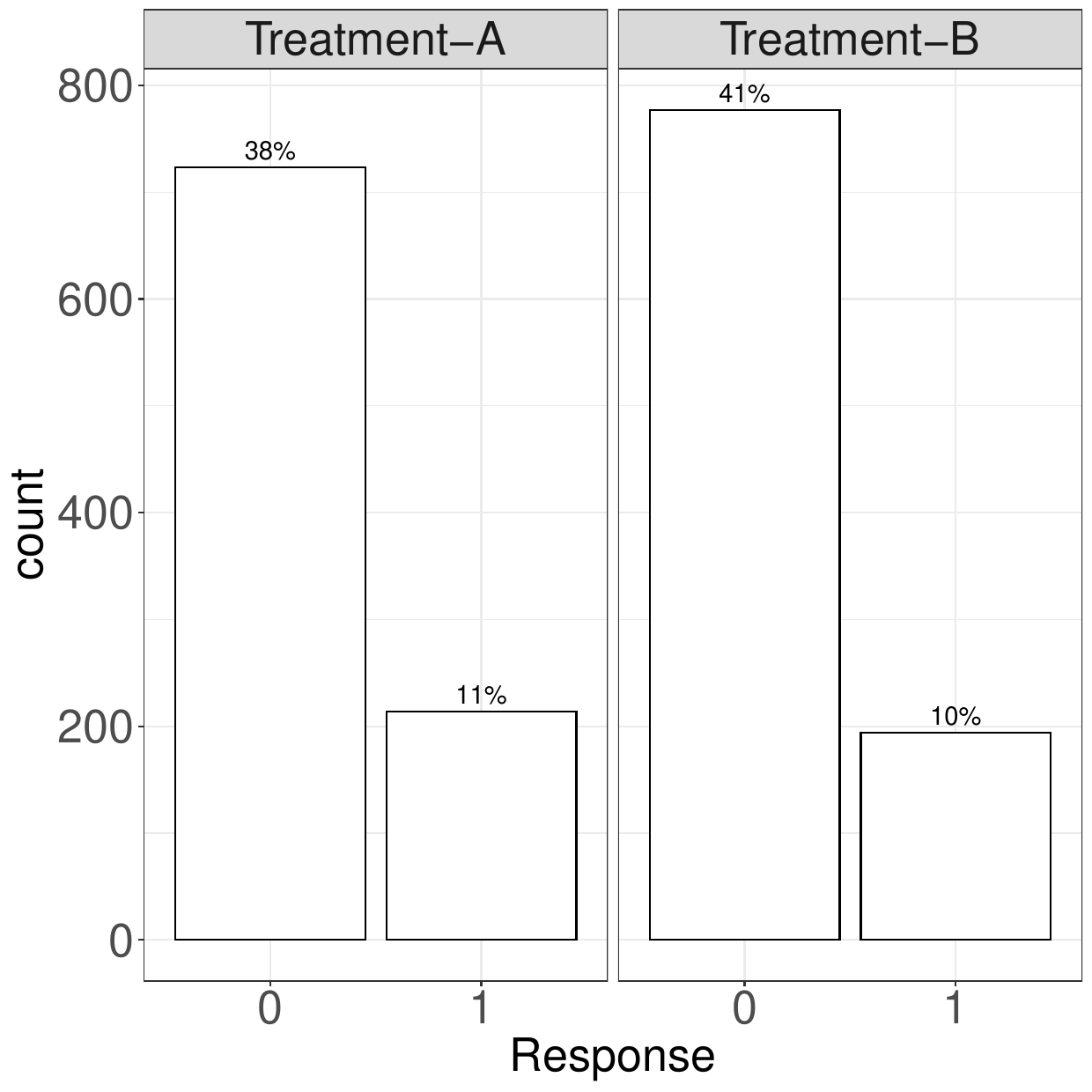}
\begin{center}
    (a)
\end{center}
\end{minipage} 
\begin{minipage}[b]{0.30\linewidth}
\includegraphics[width=\linewidth]{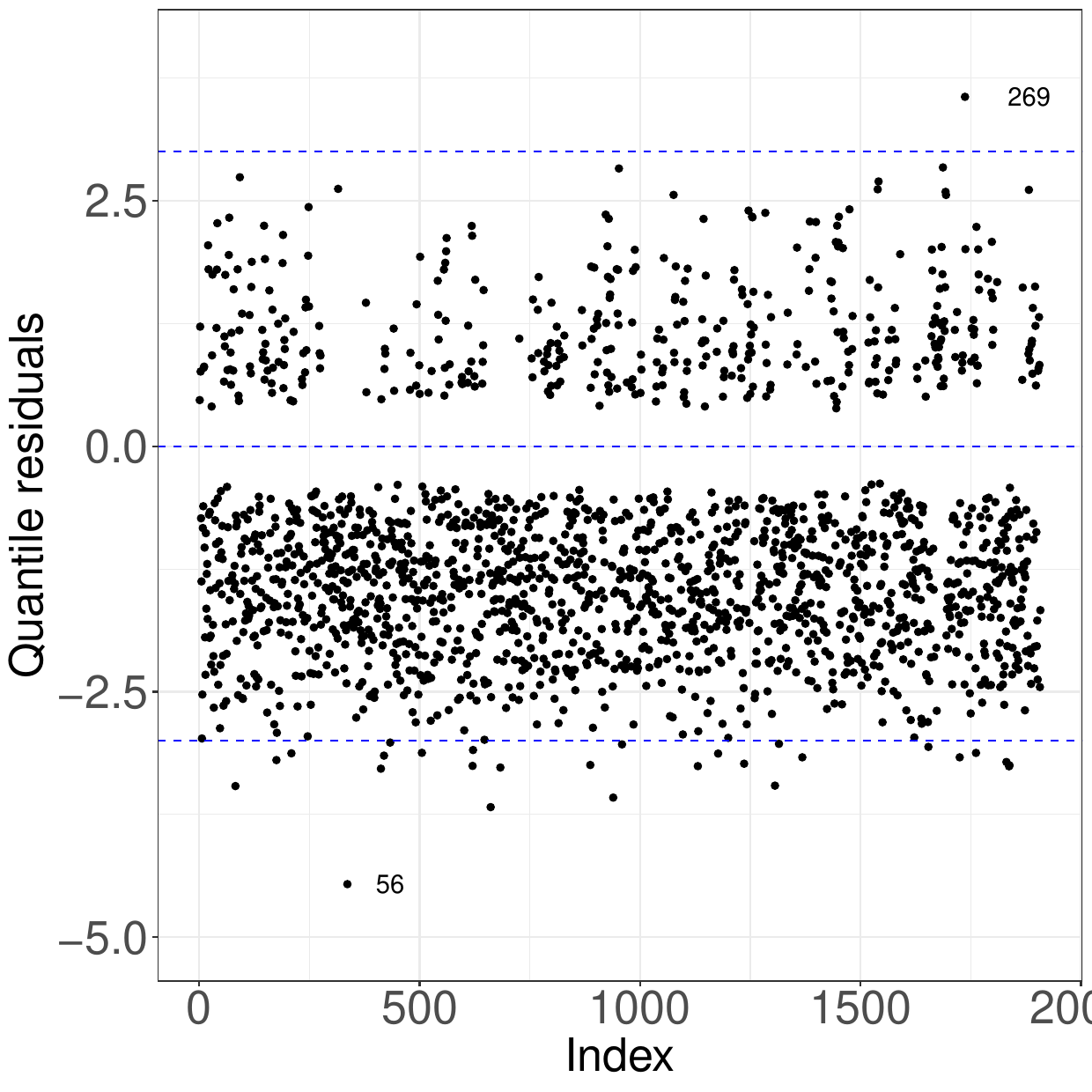}
\begin{center}
    (b)
\end{center}
\end{minipage} 
\begin{minipage}[b]{0.30\linewidth}
\includegraphics[width=\linewidth]{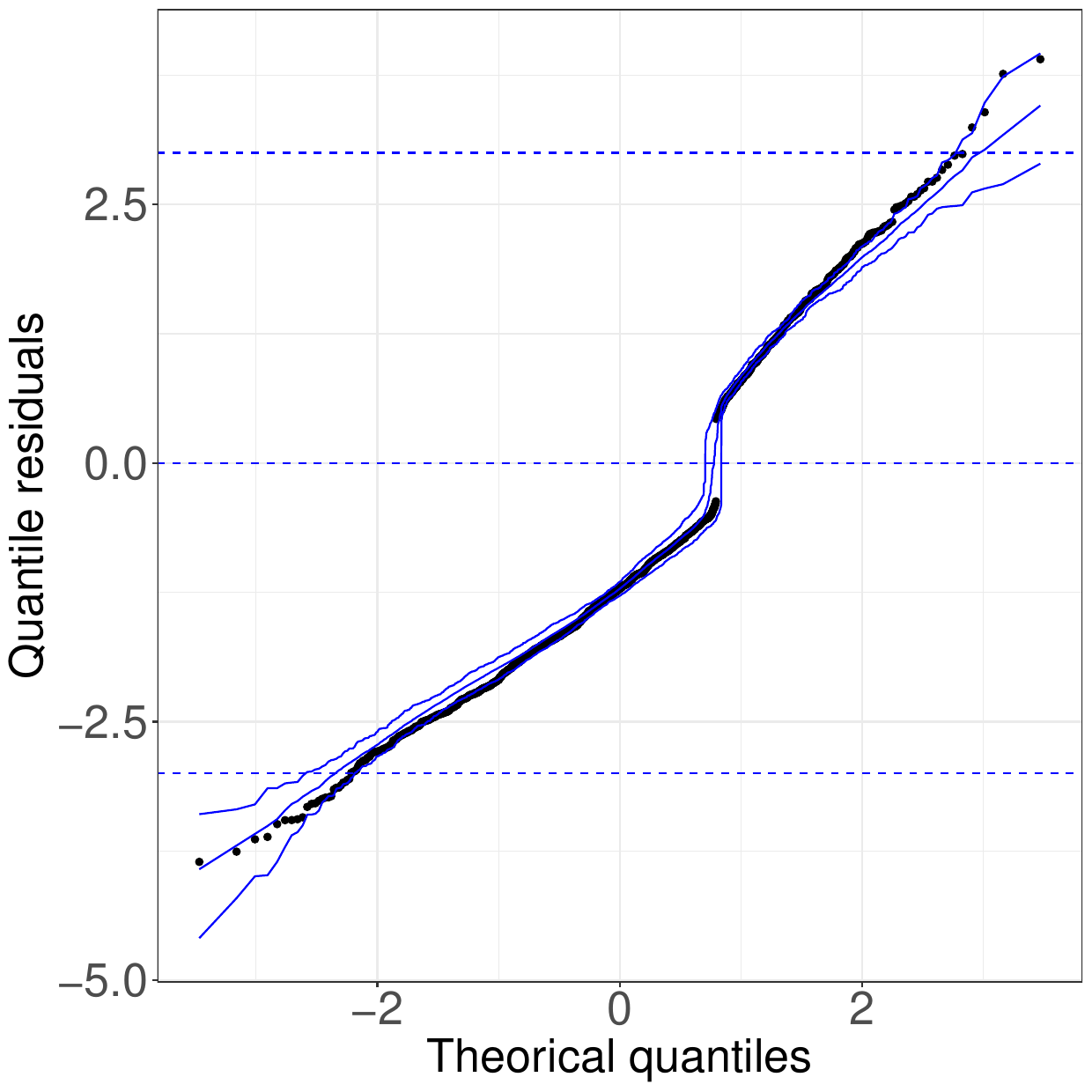}
\begin{center}
    (c)
\end{center}
\end{minipage}
\caption{Bar plot (panel left), quantile residuals versus index observations (panel center),  and simulation envelope (panel right) for the Toenail data.}
\label{apli2-plot1}
\end{figure}

The Toenail data discussed in \cite{Geert2010} were obtained from a randomized, double-blind, parallel-group, multicenter study for the comparison of two oral treatments (coded as A and B) for Toenail Dermatophyte Onychomycosis (TDO). The present study aimed to compare the efficacy of  12 weeks of continuous therapy with Treatment A or B. In total, 378 patients, distributed over 36 centers, were randomized. Subjects were followed during 12 weeks (3 months) of treatment and followed further up to 48 weeks (12 months).  Measurements were taken at baseline every month during treatment and every 3 months afterward, resulting in a maximum of 7 measurements per subject. We will restrict our analysis to 148 patients of group B, for which the target nail was one of the two big toenails for comparison with the placebo group. In this study, the binary response is the severity of infection, 0 (not severe) and 1 (severe). Figure \ref{groupToenail} gives us evidence that the percentage of severe infections decreases over time and that these percentages do not differ between the groups.
Figure~\ref{apli2-plot1}(a) reveals that no severe binary response occurred more frequently. Based on these facts,  we proposed the following hierarchical structure to fit the Toenail data
\begin{eqnarray*}
&(i)&  \bm{y}_i \sim \text{BerGLG}(\bm{\mu}_i, \phi)\\
&(ii)&  \log(-\log(1 - \mu_{ij})) = \beta_0 + \beta_1 ({\rm Treat})_{i1} + \beta_2 ({\rm Time})_{ij2}+ \beta_3 ({\rm Treat} \times {\rm Time}) , 
\end{eqnarray*}
were $\bm{y}_i = (y_{i1}, \ldots, y_{im_i})^{\top},$ is the vector containing the binary responses of the $i$th subject over time, ${\rm Treat}_{i}$ is the treatment indicator of the $i$th subject,  ${\rm Time}_{ij}$ is the time-point at which the $j$th measurement is taken for the  $i$th subject, for  $i=1,\ldots, 394$ and $j=1,\ldots,7.$ 
Table~\ref{tab1-apli} presents the parameter estimates and their respective standard errors and $p$-values of Wald-type tests from the MBerGLGR model, random intercept Bernoulli-normal models by assuming Probit and complement log-log link functions, respectively. The proposal model fitted better to data, where the effect of the variables Time and Treat$\times$Time are more significant. The Odds Ratio suggested that $100\%(1-\exp(-0.290))=25.17\%$ and $100\%(1-\exp(-0.108-0.290))=32.83\%$, decrease in the odds of the severity of infection when a one-week increase and when there is an interaction treatment and time-point, respectively. These results are in concordance with the descriptive analysis. The parameter estimate $\lambda$ is associated with the rate of ones present in the variable response and the value $\hat{\phi} = \hat{\lambda}^{-2}=0.161$ with the precision parameter.
Figures~\ref{apli2-plot1}(b) and (c) exhibit the graphic of the randomized quantile residual and the simulation envelope based on this residual for the  $j$th measurement of $i$th subject. We can observe the atypical observations. However, there are no departures from the MBerGLGR fitted and the data. Thus, we can conclude that the average probability of treatment A to be efficient for TDO increases over time.
\begin{table}[h!]
\centering{
\caption{Estimates, standard errors for the Toenail data}
\label{tab1-apli}
\renewcommand{\arraystretch}{1.2}
\begin{tabular}{lrlrlrl}
\hline
 & \multicolumn{2}{c}{\multirow{1}{*}{Bernoulli-GLG model}} & \multicolumn{2}{c}{\multirow{1}{*}{Bernoulli-Normal}} & \multicolumn{2}{c}{\multirow{1}{*}{Bernoulli-Normal}} \\
 & \multicolumn{2}{c}{\multirow{1}{*}{$g(\mu)=\log(-\log(1-\mu))$}} & \multicolumn{2}{c}{\multirow{1}{*}{$g(\mu)=\Phi(\mu)$}} & \multicolumn{2}{c}{\multirow{1}{*}{$g(\mu)=\log(-\log(1-\mu))$}} \\ 
\hline
Parameter & Estimate & Std. error & Estimate & Std. error & Estimate & Std. error \\
\hline
$\lambda$                &    $2.495$   & $0.167$       &   $2.120$ & $-$ & $3.357$ & $-$     \\
Interc.                  &    $0.930$   & $0.733$       &   $-0.919$ & $0.229^{*}$ & $-2.020$ & $0.424^{*}$     \\
Treat.                   &    $-0.057$  & $0.450$       &   $-0.079$ & $0.308$ & $-0.133$ & $0.503$     \\
Time                     &    $-0.290$  & $0.032^{*}$   &   $-0.196$ & $0.021^{*}$ & $-0.343$ & $0.040^{*}$     \\
Treat. $\times$ Time     &    $-0.108$  & $0.050^{*}$   &   $-0.077$ & $0.033^{*}$ & $-0.125$ & $0.060^{*}$     \\
\hline
AIC           & \multicolumn{2}{c}{\multirow{1}{*}{$1238.9$}} &  \multicolumn{2}{c}{\multirow{1}{*}{$1281.9$}} & \multicolumn{2}{c}{\multirow{1}{*}{$1255.3$}}  \\
\hline
\multicolumn{7}{l}{\multirow{1}{*}{$*$: Signif. variable}}
\end{tabular}
}
\end{table}

\section{Conclusions}
\label{sec:conclu}
In this paper, we derive a new discrete multivariate distribution from a Bernoulli mixed model by assuming the GLG distribution for the random intercept within a marginal approach. We demonstrate a strong conjugacy between the variable response and random effect distributions by considering the complementary log-log link function. We obtained analytical expressions for its moments and correlation structure. The maximum likelihood estimates are computed using the quasi-Newton nonlinear optimization algorithm implemented in the {\tt optim} function in the {\tt R} software. Estimation equations based on score functions are calculated with respect to parameters as an alternative to the inferential process, as well as the expressions of the components of the observed Fisher information to obtain interval estimation and hypothesis tests on the model parameters. Monte Carlo simulation studies reveal that the regression coefficients and the $\lambda$ parameter of the MBerGLGR model are asymptotically consistent for both unbalanced and balanced responses when the random effect distribution is not misspecified. The data sets were useful in demonstrating that the MBerGLGR model is an alternative for fitting correlated response binary data. The precision parameter $\phi$ is inversely proportional to the shape parameter $\lambda$, i.e. $\phi=1/\lambda^{2}$. Randomized quantile residuals for correlated outcomes were suggested to check possible data departures from the MBerGLGR model and identify influential subjects and observations. 

\bibliography{biblio}   

\end{document}